\documentclass[fleqn,usenatbib]{mnras}

\usepackage{newtxtext,newtxmath}

\usepackage[T1]{fontenc}
\usepackage{ae,aecompl}

\usepackage{graphicx}	
\usepackage{amsmath}	
\usepackage{newtxtext,newtxmath}    
\usepackage{threeparttable, tablefootnote}

\usepackage{amsmath, epsfig,natbib}
\usepackage{multicol}
\usepackage{color,ulem}
\usepackage{newtxtext,newtxmath}
\definecolor{webgreen}{rgb}{0,.5,0}
\definecolor{webbrown}{rgb}{.6,0,0}

\newcommand{\kms}{\mbox{$\>{\rm km\, s^{-1}}$}}
\newcommand{\pc}{\>{\rm pc}}
\newcommand{\kpc}{\mbox{$\>{\rm kpc}$}} 

\newcommand{\Gyr}{\mbox{$\>{\rm Gyr}$}}
\newcommand{\Myr}{\mbox{$\>{\rm Myr}$}}

\newcommand\degrees{^\circ}
\newcommand\gaia{{\it Gaia}}
\newcommand{\avg}[1]{\mbox{$\left<{#1}\right>$}}

\newcommand{\vb}{\mbox{$V_{\rm breath}$}}

\DeclareMathOperator{\sech}{sech}

\title[Breathing motion by tidally-induced spiral]{Excitation of vertical breathing motion in disc galaxies by tidally-induced spirals in fly-by interactions}

\author[A. Kumar et al.]{
Ankit Kumar,$^{1,2}$\thanks{E-mail: ankit4physics@gmail.com (AK)}
Soumavo Ghosh,$^{3}$\thanks{E-mail: ghosh@mpia.de (SG)}
Sandeep Kumar Kataria,$^{4}$
Mousumi Das$^{1}$,
Victor P. Debattista$^5$
\\
$^{1}$Indian Institute of Astrophysics, Bengaluru, 560034, India\\
$^{2}$Joint Astronomy Program, Department of Physics, Indian Institute of Science, Bengaluru, 560012, India \\
$^{3}$Max-Planck-Institut f\"{u}r Astronomie, K\"{o}nigstuhl 17, D-69117 Heidelberg, Germany\\
$^{4}$ School of Physics and Astronomy, Shanghai Jiao Tong University, No.800, Dongchuan Road, Minhang District, Shanghai, 200240, China\\
$^5$Jeremiah Horrocks Institute, University of Central Lancashire, Preston PR1 2HE, UK\\
}

\date{Accepted XXX. Received YYY; in original form ZZZ}

\pubyear{2022}

\begin{document}
\label{firstpage}
\pagerange{\pageref{firstpage}--\pageref{lastpage}}
\maketitle

\begin{abstract}
It is now clear that the stars in the Solar neighbourhood display large-scale coherent vertical breathing motions. At the same time, Milky Way-like galaxies experience tidal interactions with satellites/companions during their evolution. While these tidal interactions can excite vertical oscillations, it is still not clear whether vertical breathing motions are excited \textit{directly} by the tidal encounters or are driven by the tidally-induced spirals. We test whether excitation of breathing motions are directly linked to tidal interactions by constructing a set of $N$-body models (with mass ratio 5:1) of unbound, single fly-by interactions with varying orbital configurations. We first reproduce the well-known result that such fly-by interactions can excite strong transient spirals (lasting for $\sim 2.9-4.2 \Gyr$) in the outer disc of the host galaxy. The generation and strength of the spirals are shown to vary with the orbital parameters (the angle of interaction, and the orbital spin vector). Furthermore, we demonstrate that our fly-by models exhibit coherent breathing motions whose amplitude increases with height. The amplitudes of breathing motions show characteristic modulation along the azimuthal direction, with compressing breathing motions coinciding with the peaks of the spirals and expanding breathing motions falling in the inter-arm regions -- a signature of a spiral-driven breathing motion. These breathing motions in our models end when the strong tidally-induced spiral arms fade away. Thus, it is the tidally-induced spirals which drive the large-scale breathing motions in our fly-by models, and the dynamical role of the tidal interaction in this context is indirect.
\end{abstract}



\section{Introduction}
\label{sec:intro}

The second data release from the \gaia\ mission (hereafter \gaia\ DR2) has revealed the presence of large-scale bulk vertical motions ($\sim 10 \kms$ in magnitude) and the associated bending and breathing motions for stars in the Solar vicinity and beyond \citep{Gaia.Collaboration.2018}. The presence of such breathing motions, i.e., stars on both sides of the Galactic mid-plane moving coherently towards or away from it, has also been reported in various past Galactic surveys, for example, the SEGUE (Sloan Extension for Galactic Understanding and Exploration) survey \citep{Widrow.etal.2012}, the LAMOST (Large Sky Area Multi-Object Fibre Spectroscopic Telescope) survey  \citep{Carlin.etal.2013}, and the RAVE (Radial Velocity Experiment) data \citep{Williams.etal.2013}. The existence of such non-zero bulk vertical motions in the Milky Way raises questions about the plausible driving mechanism(s), since, in an axisymmetric potential, the bulk radial and vertical motions should be zero \citep[e.g.][]{BT08}.
\par
Much of the understanding of the excitation of breathing motions in Milky Way-like galaxies have been gleaned from numerical simulations. Using semi-analytic models and test-particle simulations, \citet{Faureetal2014} was the first to show that a strong spiral can drive large-scale vertical motions ($|\avg{v_z}| \sim 5-20 \kms$). The amplitude of such breathing motion increases at first with height from the mid-plane, and then starts to decrease after reaching its maximum value at a certain height. Also, using self-consistent $N$-body simulation, \citet{Debattista.2014} showed that a \textit{vertically-extended} spiral feature can drive strong large-scale breathing motions, with amplitude increasing monotonically from the mid-plane. The relative sense of these bulk motions, whether compressing or expanding, changes across the corotation resonance (hereafter CR) of the spiral \citep{Faureetal2014,Debattista.2014}. Furthermore, using a {\it self-consistent}, high-resolution simulation with star formation,  \citet{Ghosh.etal.2020} studied the age-dependence of such vertical breathing motions excited by spiral density waves. They showed that, at fixed height, the amplitude of such vertical breathing motion decreases with stellar age. They showed a similar age-variation in the breathing amplitude in the \gaia\ DR2, thereby supporting the scenario that the breathing motion of the Milky Way might well be driven by spiral density waves \citep{Ghosh.etal.2020}. Instead, \citet{Monari.etal.2015} showed that a stellar bar can also drive such breathing motion in disc galaxies. However, the resulting amplitudes of the breathing motions are small ($|\avg{v_z}| \sim 1 \kms$) when compared to the spiral-driven breathing amplitudes. 
\par
In the Lambda cold dark matter ($\Lambda$CDM) paradigm of hierarchical structure formation, galaxies grow in mass and size via major mergers and/or multiple minor mergers, and cold gas accretion \citep{WhiteandRees1978,Fall1980}. During the evolutionary phase, a galaxy also experiences multiple tidal interactions with satellites and/or passing-by companion galaxies. The frequency of such fly-by encounters increases at lower redshifts \citep[e.g., see][]{SinhaandBockelmann2015}, and their cumulative dynamical impact on the morphology as well as on the dynamics of the host galaxies can be non-negligible \citep[e.g., see][]{Anetal2019}.
 Fly-by encounters can excite an $m=2$ bar mode \citep[e.g., see][]{Noguchi1987,Lokasetal2016,Martinez-Valpuestaetal2017,Ghoshetal2021a}, off-set bars with a one-arm spiral \citep[e.g,][]{Pardyetal2016}, and an $m=1$ lopsidedness in the stellar disc \citep{Bournaudetal2005,Mapellietal2008,Ghoshetal2021b}. They can also trigger star formation \citep[e.g.,][]{Ducetal2018}, and if  the gas inflow towards the centre is exceedingly large, this can lead to a starburst \citep{MihosandHernquist1994} as well as triggering of AGN activity \citep{Combes2001}. Furthermore, the role of fly-bys has been investigated in the context of forming warps \citep{Kimetal2014,Semczuketal2020}, disc heating and disc thickening \citep{ReshetnikovandCombes1997, Kumar.etal.2021}, tidal bridges and streams \citep[e.g., see][]{ToomreToomre1972,DucandRenaud2013}, altering the galaxy spin \citep{ChoiandYi2017}, and the evolution of classical and pseudo-bulges \citep{Kumar.etal.2021}. 
 Our Galaxy has also experienced such a tidal interaction with the Sagittarius (Sgr) dwarf galaxy \citep[e.g., see][]{Majewskietal2003}. Recent studies have indicated that such a tidal interaction could excite a `snail-shell' structure (phase-space spiral), bending motions in the Solar neighbourhood \citep[for details see, e.g.,][]{Widrowetal2014,Antojaetal2018,Gaia.Collaboration.2018}. 
\par
A tidal encounter with another galaxy can excite spiral features, as was first proposed in the seminal work of \citet{Holmberg1941}. Later, pioneering  numerical work of \citet{ToomreToomre1972} showed that a tidal interaction can excite tidal tails, bridges, and spiral features in the disc of the host galaxy for a wide variety of orbital configurations. Following that, numerical simulation has become an indispensable tool to study the dynamical effect of galaxy interactions. Several past studies attempted to understand the role of tidal encounters in the context of excitation of spirals as well as to understand the longevity and nature of the resulting spirals \citep[e.g. see][]{Sundeliusetal1987,DonnerandThomasson1994,SaloandLaurikainen2000,dobbs.etal.2010,Pettittetal2018}. 
Furthermore, a recent study by \citet{Pettittetal2017} investigated star formation and the properties of interstellar medium in tidally-induced spirals. 
\par
Tidal interactions can also induce vertical distortions and oscillations in the disc of the host galaxy \citep[e.g., see][]{HunterandToomre1969,Araki1985,Mathur1990,Weinberg1991,VesperiniandWeinberg2000,Gomezetal2013,Widrowetal2014,Donghiaetal2016}. \citet{Widrowetal2014} proposed a dynamical scenario where a satellite galaxy, while plunging into the disc, can excite both bending and breathing motions. Interestingly, such tidal  interactions also excited a strong spiral response within the disc in their model (see Fig.~10 there). Therefore, it is still unclear whether the tidal interactions are `directly' responsible for driving breathing motion, or the tidally-induced spirals are driving the breathing motions.
\par
 We aim to test this latter hypothesis in detail in this paper. We study a set of $N$-body models of galaxy fly-by interactions while varying the orbital parameters. We investigate the generation of the spiral features due to such fly-by encounters, and quantify the nature and longevity of such spirals in different fly-by models. We closely follow the generation of the vertical breathing motions and their subsequent evolution. In particular, we look for evidence that the generation and evolution of the vertical breathing motions is correlated with the temporal evolution of the tidally-induced spirals.
\par
 The rest of the paper is organized as follows. Section~\ref{sec:sim_setup} provides the details of the simulation set-up and the fly-by models. Section~\ref{sec:spiral_quant} presents the results of the tidally-induced spirals, the density wave nature of spirals, and the temporal evolution of their strength. Section~\ref{sec:breathing} measures the properties of the vertical breathing motions driven by these tidally-induced spirals. Section~\ref{sec:discussion} discusses a few limitations of this work while section \ref{sec:conclusion} summarizes the main findings of this work.  

\section{Simulation set-up of galaxy Fly-by models}
\label{sec:sim_setup}

\begin{table*}
\centering
\caption{Key galaxy parameters for the equilibrium models of the host and the perturber galaxies.}
\begin{tabular}{ccccccccccccc}
\hline
Galaxy & $M$$^{(1)}$ & $\lambda$$^{(2)}$ & $c$$^{(3)}$ & $f_{\rm disc}$$^{(4)}$ & $f_{\rm bulge}$$^{(5)}$ & $j_{\rm d}$$^{(6)}$ &  $R_{\rm d}$$^{(7)}$ & $z_0$$^{(8)}$ & $N_{\rm halo}$$^{(9)}$ & $N_{\rm disc}$$^{(10)}$ & $N_{\rm bulge}$$^{(11)}$ & $N_{\rm tot}$ $^{(12)}$\\
& ($\times 10^{12} M_{\odot}$) &&&&&& ($\kpc$) & ($\kpc$) & ($\times 10^{6}$) & ($\times 10^{6}$) & ($\times 10^{6}$) & ($\times 10^{6}$)\\
\hline
Host & 1.2 & 0.035 & 10 & 0.025 & 0.005 & 0.03 & 3.8 & 0.38 & 2.5 & 1.5 & 1 & 5 \\
Perturber & 0.24 & 0.035 & 11 & 0.01 & 0.002 & 0.01 & 1.95 & 0.195 & 0.5 & 0.3 & 0.2 & 1\\
\hline
\end{tabular}
\centering
{ \hspace{0.9 cm} (1) total mass (in $M_{\odot}$); (2) halo spin; (3) halo concentration; (4) disc mass fraction; (5)  bulge mass fraction; (6) disc spin fraction; (7)  disc scale length (in $\kpc$); (8) disc scale height (in $\kpc$); (9) total DM halo particles; (10) total disc particles; (11) total bulge particles; (12) total number of particles used. }
\label{tab:initial_parameters}
\end{table*}

To motivate our study, we construct a set of $N$-body models of galaxy fly-bys where the host galaxy experiences an unbound interaction with a perturber galaxy. The mass ratio of the perturber and the host galaxy is set to 5:1, and is kept fixed for all the models considered here. A prototype of such a galaxy fly-by model is already presented in  \citet{Kumar.etal.2021}. Here, we construct a suite of fly-by models varying the orbital configuration (e.g., angle of interaction, orientation of the orbital spin vector). The details of the modelling and the simulation setup is discussed in  \citet{Kumar.etal.2021}. For the sake of completeness, here we briefly mention the equilibrium model of the galaxies as well as the orbital configurations of the galaxy interaction.

\subsection{Equilibrium models}
\label{sec:equilibrium_model}

The initial equilibrium model of each galaxy (host and the perturber) consists of a classical bulge, a stellar disc, and a dark matter (hereafter DM) halo. Each of the galactic components is treated as live, thereby allowing them to interact with each other. The DM halo is assumed to be spherically symmetric, and is modelled with a Hernquist density profile \citep{Hernquist1990} of the form 
\begin{equation}
    \rho_{\rm dm}(r)=\frac{M_{\rm dm}}{2\pi}\frac{a}{r(r+a)^3}\,,
    \label{eqn:halo}
\end{equation}
\noindent where $M_{\rm dm}$ and $a$ are the total mass and the scale radius of the DM halo, respectively. The scale radius of the Hernquist halo is related to the concentration parameter `$c$' of NFW DM halo \citep{NFW1996}. For an NFW DM halo with mass $M_{200}=M_{\rm dm}$, this relation is given by the following equation,
\begin{equation}
    a=\frac{r_{200}}{c}\sqrt{2\left[\ln{(1+c)-\frac{c}{(1+c)}}\right]},
    \label{eqn:scale_radius}
\end{equation}
where $r_{200}$ represents the radius of an NFW halo\footnote{It is defined as the radius from the centre of the halo inside which the mean density is 200 times the critical density of the Universe.}, and the mass within this radius is defined as $M_{200}$. The classical bulge is also assumed to be spherically symmetric and is modelled with another Hernquist density profile \citep{Hernquist1990} of the form
\begin{equation}
    \rho_{\rm b}(r)=\frac{M_{\rm b}}{2\pi}\frac{b}{r(r+b)^{3}}\,,
    \label{eqn:bulge}
\end{equation}
where $M_{\rm b}$ and $b$ represent the total bulge mass and the bulge scale radius, respectively.  The initial radial surface density of the stellar disc follows an exponential fall-off and has a $\mathrm {sech}^2$ profile along the vertical direction, thereby having the form 
\begin{equation}
   \rho_{d}(\rm R,\rm z)=\frac{M_{\rm d}}{4\pi z_{0} R_{\rm d}^{2}}\exp\left(-{\frac{R}{R_{\rm d}}}\right) \sech^{2}\left(\frac{z}{z_{0}}\right), 
    \label{eqn:disc}
\end{equation}
\noindent where $M_{\rm d}$ is the total mass, $R_{\rm d}$ is the exponential disc scale length, and  $z_{0}$ is the scale height. The corresponding values of the structural parameters, used to model the host as well as the perturber galaxy, are listed in Table~\ref{tab:initial_parameters}.
\par
\begin{figure*}
  \begin{multicols}{2}
	\includegraphics[width=0.95\linewidth]{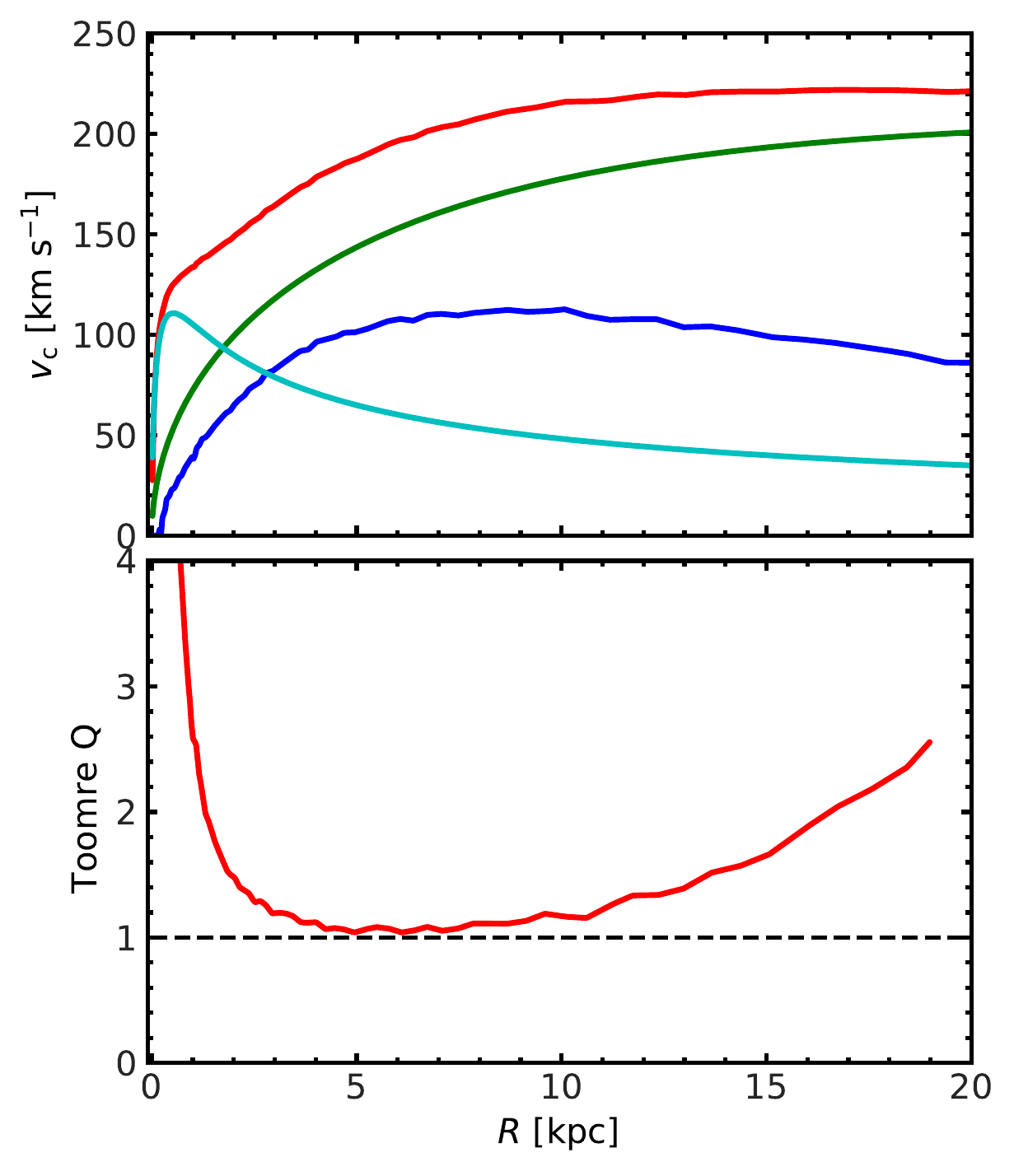}\par
	\includegraphics[width=0.95\linewidth]{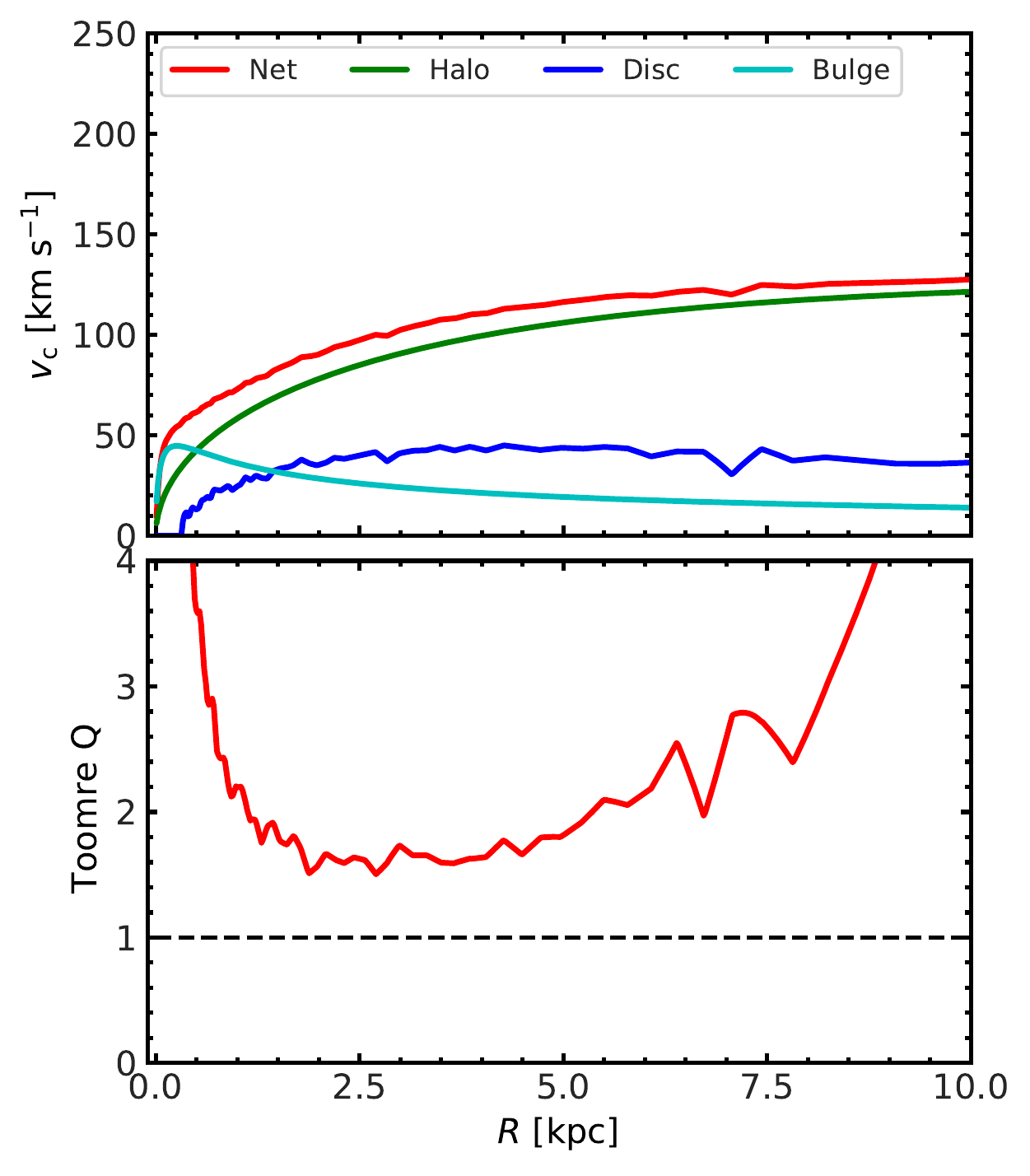}\par
	\end{multicols}
    \caption{Radial profiles of the circular velocity ($v_{\rm c}$) and the Toomre $Q$ parameter are shown for equilibrium models of the host (left-hand panels) and the perturber galaxy (right-hand panels). In the top panels, the blue line denotes the contribution of the stellar disc while the DM halo contribution is shown by the green line. The bulge contribution is shown by the cyan solid line whereas the red line denotes the total/net circular velocity.}
    \label{fig:equilibrium_dynamics}
\end{figure*}

The equilibrium models for the host as well as the perturber are generated using the publicly distributed code {\sc galic} \citep{Yurin2014}. This code uses elements of the Schwarzschild's method and the made-to-measure method to search for a stable solution of the collisionless Boltzmann Equation (CBE) for a set of collisionless stellar particles, initialized by some predefined analytic density distribution functions \citep[for details, see][]{Yurin2014}. A total of $5 \times 10^6$ particles are used to model the host galaxy whereas a total of $1 \times 10^6$ particles are used to model the perturber galaxy. The number of particles used to model each of the galaxy components of the host and the perturber galaxy are also listed in Table~\ref{tab:initial_parameters}. The stellar particles have gravitational softening $\epsilon = 20 \pc$ while the DM halo particles have $\epsilon = 30 \pc$. Fig.~\ref{fig:equilibrium_dynamics} shows the corresponding radial profiles of the circular velocity ($v_{\rm c}$) and the Toomre $Q$ parameter for the host and the perturber galaxy at $t =0$.

\subsection{Set-up of galaxy fly-by scenario}
\label{sec:flyby_setup}

\begin{table}
\centering
\caption{Key orbital parameters for the galaxy fly-by models.}
\label{tab:sims_name}
\begin{threeparttable}
\begin{tabular}{lccccr}
\hline
Model\tnote{(a)} & $r_{\rm p}\tnote{(b)}$ & $i^{\degrees}\tnote{(c)}$ & $t_{\rm p}\tnote{(\rm d)}$ & $t_{\rm enc}\tnote{(e)}$ & $T_{\rm p}\tnote{(f)}$ \\
& (kpc) & & (Gyr) & (Gyr)\\
\hline
$RP40i00pro$ & 53.09 & 0 & 0.88 & 0.1298 & -4.1348\\
$RP40i30pro$ & 52.34 & 30 & 0.85 & 0.1272 & -4.1162\\
$RP40i60pro$ & 52.14 & 60 & 0.85 & 0.1265 & -4.1113\\
$RP40i90pro$ & 52.23 & 90 & 0.85 & 0.1267 & -4.1133\\
$RP40i00ret$ & 53.07 & 0 & 0.85 & 0.1298 & -4.1342\\
\hline
\end{tabular}
\centering
{(a) Galaxy fly-by model; (b) pericentre distance (in kpc); (c) orbital angle of interaction (in degree); (d) time of pericentre passage (in Gyr); (e) encounter time (in Gyr); (f) tidal parameter.}

\end{threeparttable}
\end{table}

To simulate the unbound galaxy fly-by scenario, we place our galaxy models on a hyperbolic orbit with eccentricity, $e=1.1$ so that the orbit of the perturber galaxy remains unbound throughout the interaction. We avoid choosing a parabolic orbit as the dynamical friction of the host galaxy decays the orbit of the perturber galaxy and puts the perturber galaxy on a bound elliptical orbit. Our choice of hyperbolic orbit avoids a bound fly-by interaction. We place the galaxies at an initial separation of $255 \kpc$ before the start of the simulation. For different models with different orbital configurations, we vary the distance of their closest approach (the pericentre distance) assuming the two-body Keplerian orbit.  
For further details of the orbital configuration and the geometry of the unbound fly-by scenario, the reader is referred to \citet{Kumar.etal.2021}. A total of five such galaxy fly-by models are used for this study. The angle of interaction, and the pericentre distance for these models are listed in Table~\ref{tab:sims_name}.
\par
All the simulations are run using the publicly available code {\sc{Gadget-2}} \citep{Springel2001, Springal2005man, Springel2005} for a total time of $6 \Gyr$, with a tolerance parameter $\theta_{\rm tol} = 0.5$ and an integration time-step $0.4 \Myr$. The maximum error in the angular momentum of the system is well within $0.15$ percent throughout the evolution for all models considered here.
\par
Following \citet{dimatteo.etal.2007}, we also calculate the encounter time ($t_{\rm enc}$), and tidal parameter ($T_{\rm p}$) for the host galaxy by using 
 \begin{equation}
    t_{\rm enc} = \frac{r_{\rm p}}{v_{\rm p}},\\
    T_{\rm p} = \log_{10} \left[\frac{m_{\rm per}}{M_{\rm host}} \left(\frac{R_{\rm d}}{r_{\rm p}} \right)^{3}\right],
\end{equation}
\noindent where $r_{\rm p}$ is pericentre distance and $v_{\rm p}$ is the corresponding relative velocity. $M_{\rm host}$ and $m_{\rm per}$ are the masses of the host and the perturber galaxy, respectively. The corresponding values of $t_{\rm enc}$, and $T_{\rm p}$ for all the models consider here, are listed in Table~\ref{tab:sims_name}. We note that the quantities mentioned here will be in units of the dimensionless parameter `$h$', defined via the Hubble constant ($H_{0}=100 h$ km s$^{-1}$ Mpc$^{-1}$) and can be scaled to observed values.
 \par
 Each model is referred as a unique string given by `{\sc [pericentre distance][angle of interaction][orbital spin]}' where {\sc [pericentre distance]} denotes the pericenter distance, obtained using the standard two-body formalism. {\sc [angle of interaction]} denotes the angle at which the perturber encounters with the host galaxy while {\sc [orbital spin]} denotes the orbital spin vector (`pro' for prograde and `ret' for retrograde orbits). We follow this scheme of nomenclature throughout the paper.  As an example, RP40i30pro denotes a fly-by model where  the perturber galaxy interacts with the host galaxy at an angle of $30 \degrees$ in a prograde orbit, and the calculated pericentre distance between these two galaxies is $40 \kpc$, obtained by using the standard two-body  Keplerian orbit.

\section{Quantification of tidally-induced spirals in galaxy fly-bys}
\label{sec:spiral_quant}
%
\begin{figure*}
    \centering
	\includegraphics[width=\textwidth]{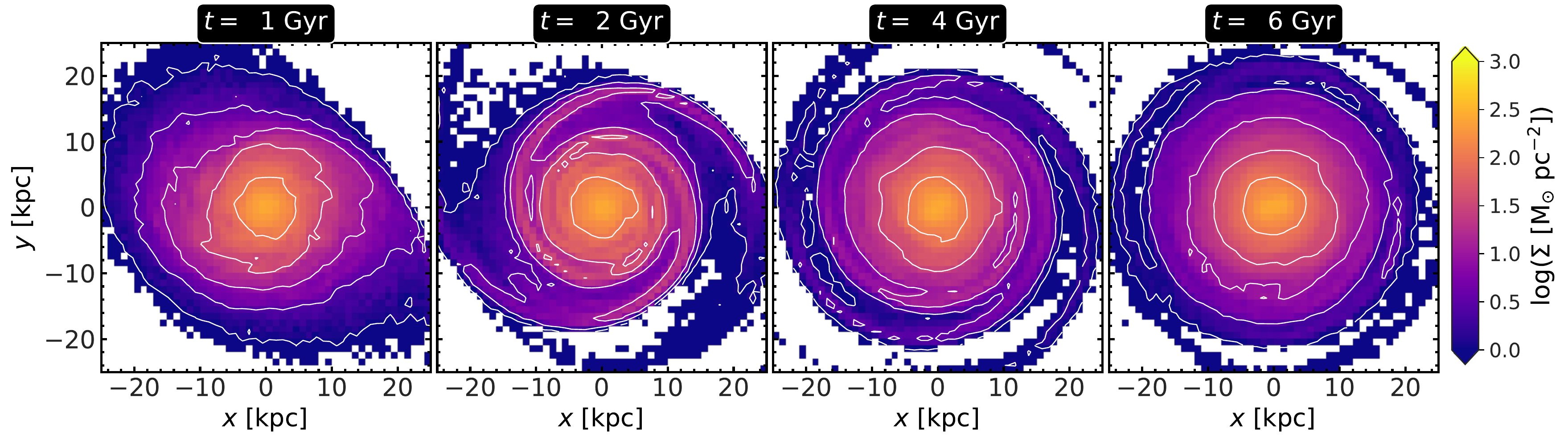}
	\includegraphics[width=\textwidth]{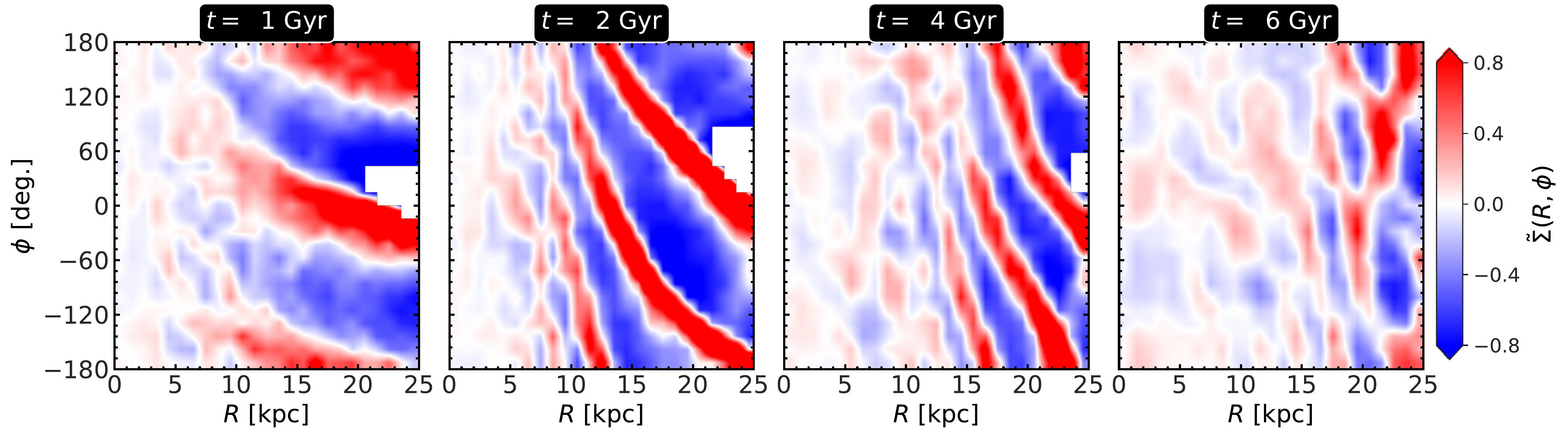}
    \caption{ Top: the density distribution of the stars of the host galaxy in the face-on projection, at four different times after the pericentre passage, for the model  $RP40i00pro$. White solid lines show the contours of constant surface density. A strong spiral feature is excited in the disc of the host galaxy after it experiences a tidal encounter with the perturber galaxy. Bottom: the corresponding distribution of the residual surface density ($\tilde \Sigma (R, \phi)$), calculated using Eq.~\ref{eqn:res_dens}, at the same four different times. The sense of rotation is towards increasing $\phi$.}
    \label{fig:dens_map}
\end{figure*}
Here, we investigate the excitation of spiral structure in the host galaxy as a consequence of the tidal interaction with the perturber galaxy. We first study the model $RP40i00pro$, where the host galaxy experiences an in-plane, unbound fly-by interaction with the perturber galaxy (with mass $1/5$th that of the host galaxy). During the pericentre passage, the perturber exerts a strong tidal pull on the host galaxy. Fig.~\ref{fig:dens_map} (top panels) shows the face-on density distribution of the stars of the host galaxy at four different times, after the tidal interaction occurs. A visual inspection reveals that after the interaction happens, a strong spiral feature is excited in the outer parts of the host galaxy (e.g., at $t = 2 \Gyr$). The spiral features are also seen at later times. However, by the end of the simulation ($t = 6 \Gyr$), there is no discernible, strong spiral features left in the host galaxy.
To study this trend further, we calculate the residual surface density ($\tilde \Sigma(R, \phi)$) in the $(R, \phi)$-plane using 
\begin{equation}
    \tilde \Sigma (R, \phi) = \frac{\Sigma (R, \phi) - \Sigma_{\rm avg} (R)}{\Sigma_{\rm avg} (R)},
    \label{eqn:res_dens}
\end{equation}
\noindent where $\Sigma_{\rm avg} (R)$ is the azimuthally-averaged surface density of the disc at radius $R$. This is shown in  Fig.~\ref{fig:dens_map} (bottom panels). As seen clearly, after the interaction happens, a strong, coherent spiral feature, denoted by the presence of a periodic over- and under-density, is excited in the outer region ($R \geq 10 \kpc$) of the host galaxy. At the end of the simulation ($t = 6 \Gyr$), the corresponding residual density distribution does not exhibit any coherent spiral structure in the disc of the host galaxy.

\subsection{Strength and temporal evolution of spirals}
\label{sec:strength_spiral}

\begin{figure}
    \centering
	\includegraphics[width=\columnwidth]{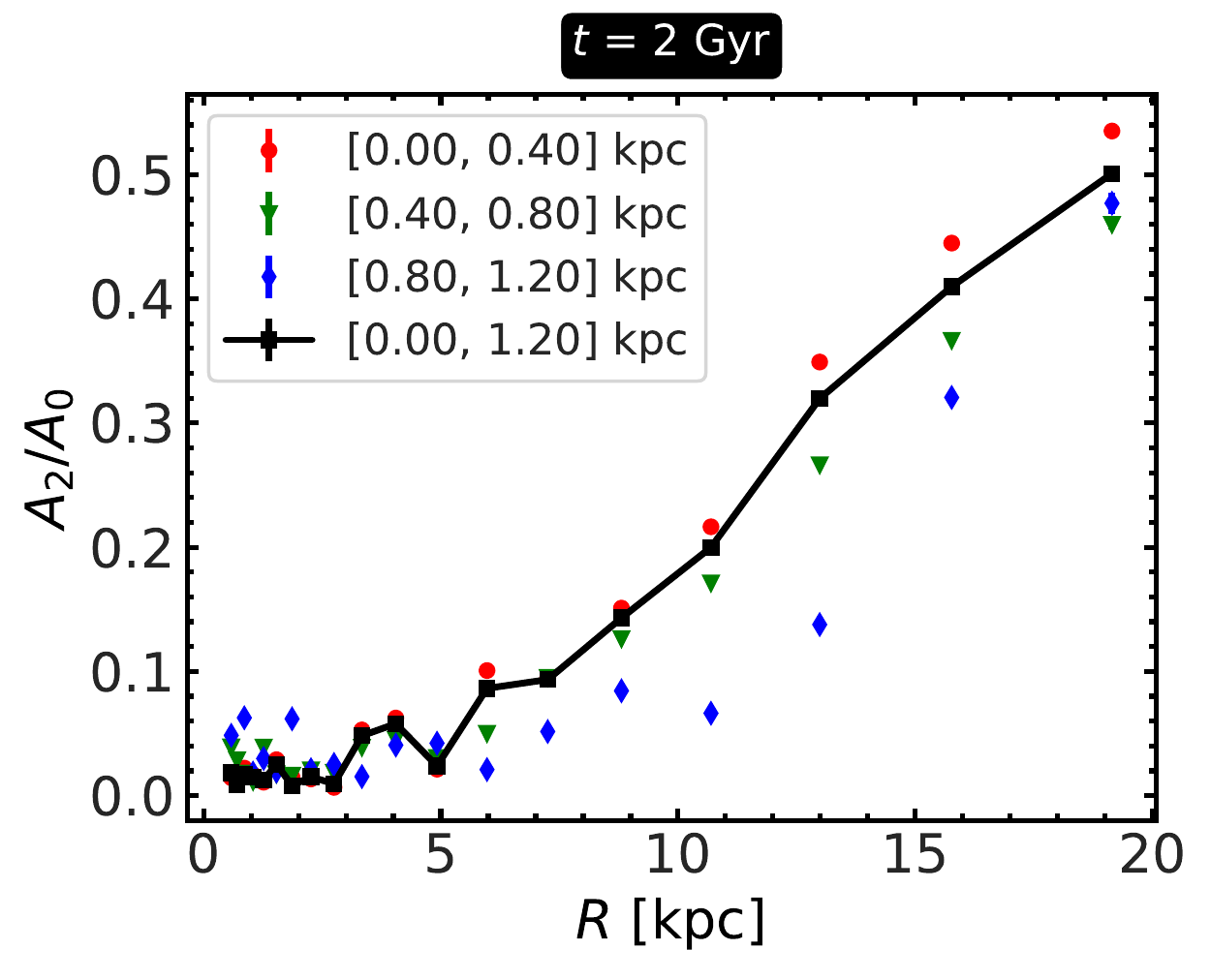}
    \caption{Radial variation of the Fourier coefficient of the $m=2$ Fourier component (normalised by the $m=0$ component) is shown at $t = 2 \Gyr$ for the model $RP40i00pro$ (see the black solid line). The same quantity is then measured for stars in different vertical layers (as indicated in the legend).}
    \label{fig:m2_strength_dz_0.4}
\end{figure}

In the previous section, we have shown that a tidal interaction with a perturber galaxy excites a prominent spiral feature in the outer disc region of the host galaxy for the model $RP40i00pro$. Next, we quantify the strength of the spiral and follow its temporal evolution. For this, we first calculate the radial variation of the Fourier moment of the surface density of the stellar particles of the host galaxy using 
\begin{align}
A_m/A_0 (R)= \left|\frac{\sum_j m_j e^{im\phi_j}}{\sum_j m_j}\right|\,,
    \label{eqn:fourier_mode}
\end{align}
\noindent where $A_m$ is the coefficient of the $m$th Fourier moment of the density distribution, $m_j$ is the mass of the $j$th particle \citep[e.g., see][]{Kumar.etal.2021, Kumar.etal.2022}. Fig.~\ref{fig:m2_strength_dz_0.4} shows the corresponding radial variation of the $m=2$ Fourier moment at $t = 2 \Gyr$ for the model $RP40i00pro$. In the outer disc regions, there are less particles when compared to the inner disc regions. Therefore, using a \textit{linearly-spaced} radial binning when calculating the Fourier coefficient (using Eq.~\ref{eqn:fourier_mode}) introduces noise in the calculation for the outer regions. To avoid that, we employ a logarithmic binning in the radial direction. As seen clearly, in the outer parts ($R \geq 10 \kpc$), the values of the coefficient $A_2/A_0$ are non-zero, indicating the presence of a strong spiral structure. Are these tidally-induced spirals mostly confined to the disc mid-plane or are they vertically-extended? To investigate this further, we calculate the radial variation of the same Fourier coefficient $A_2/A_0$, but for stars in different vertical layers of thickness $400 \pc$. The resulting radial variations are also shown in Fig.~\ref{fig:m2_strength_dz_0.4}. As seen clearly from Fig.~\ref{fig:m2_strength_dz_0.4}, the Fourier coefficient $A_2/A_0$ shows non-zero values even for stars at the largest heights from the mid-plane ($|z| = [0.8, 1.2] \kpc$). Also, at a certain radius $R$ within the extent of the spirals, the values of the Fourier coefficient $A_2/A_0$ decreases monotonically as one moves farther away from the mid-plane. We checked this variation of the Fourier coefficient $A_2/A_0$ with height at other time-steps as well, and found that this trend remains generic whenever the tidally-induced spirals are strong in the disc of the host galaxy. This demonstrates that the tidally-induced spirals in our model $RP40i00pro$ is \textit{vertically-extended}, similar to what was reported in \citet{Debattista.2014} and \citet{Ghosh.etal.2020}. 
\par
\begin{figure}
    \centering
	\includegraphics[width=1.03\linewidth]{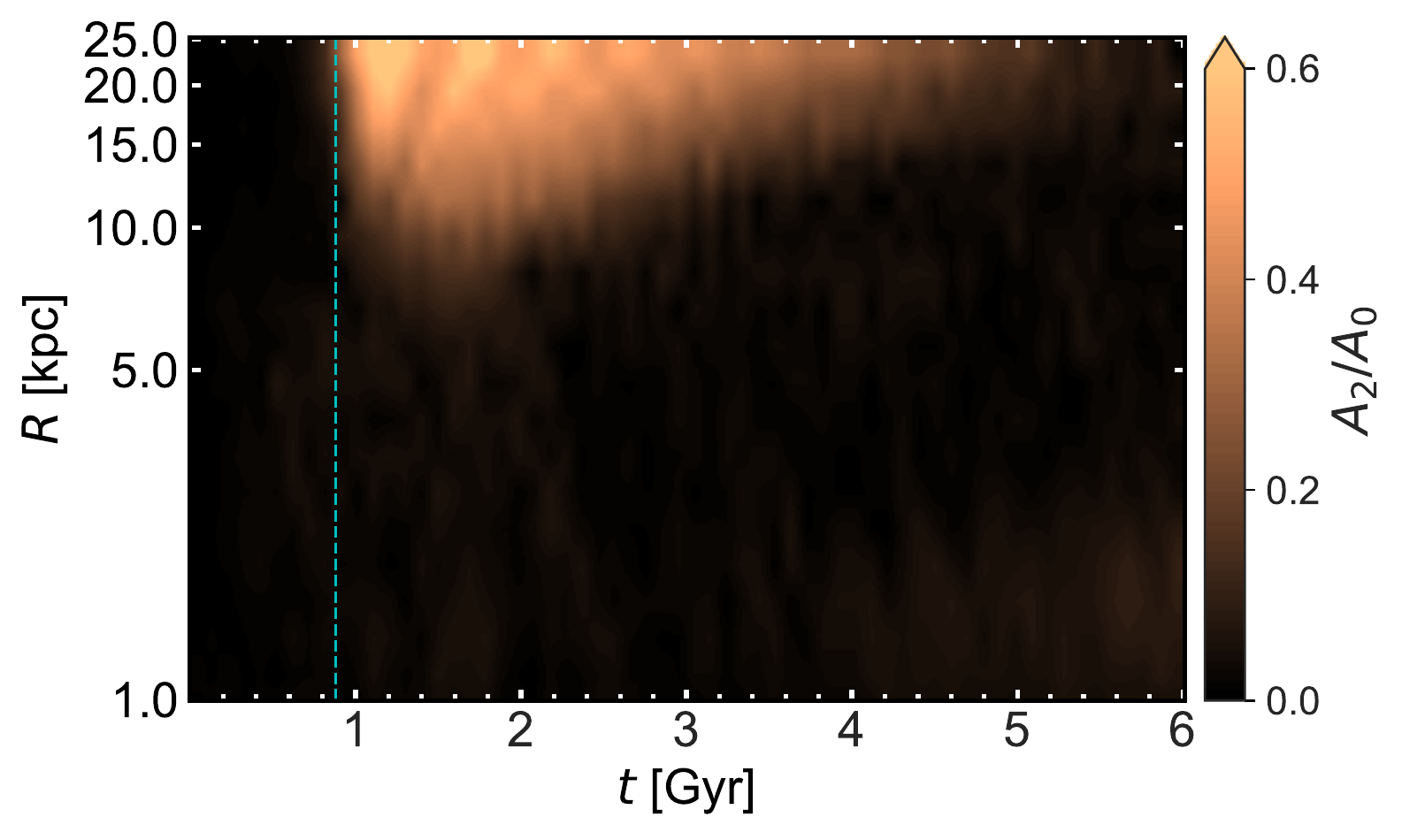}
    \caption{The evolution of the coefficient of the $m=2$ Fourier moment ($A_2/A_0$) in the $R-t$ space for the model $RP40i00pro$. A logarithmic binning is employed along the radial direction, for details see text. The colour bar shows the values of $A_2/A_0$. The vertical dashed line (in cyan) denotes the time of the pericentre passage of the perturber.}
    \label{fig:spiral_strength_t-R}
\end{figure}

To further investigate the spatio-temporal evolution of the tidally-induced spirals in model $RP40i00pro$, we calculate the Fourier coefficient $A_2/A_0 (R)$ at different radial locations for the whole simulation run-time (a total of $6 \Gyr$). This is shown in Fig.~\ref{fig:spiral_strength_t-R}. The tidal interaction with the perturber excites a strong spiral feature in the disc of the host galaxy after $t \sim 1.1 \Gyr$ or so. These spirals remain mostly in the outer regions of the host's disc ($R \geq 10 \kpc$), as the values of $A_2/A_0$ in the inner part ($R \leq 10 \kpc$) are almost zero (see Fig.~\ref{fig:spiral_strength_t-R}). After $t \sim 3 \Gyr$ or so, the spiral starts to decay, as shown by the decreasing values of the $A_2/A_0$. By the end of the simulation run, the values of the $A_2/A_0$ become almost zero, implying that the tidally-induced spirals have wound up almost completely. 
\par
\begin{figure}
    \centering
	\includegraphics[width=\columnwidth]{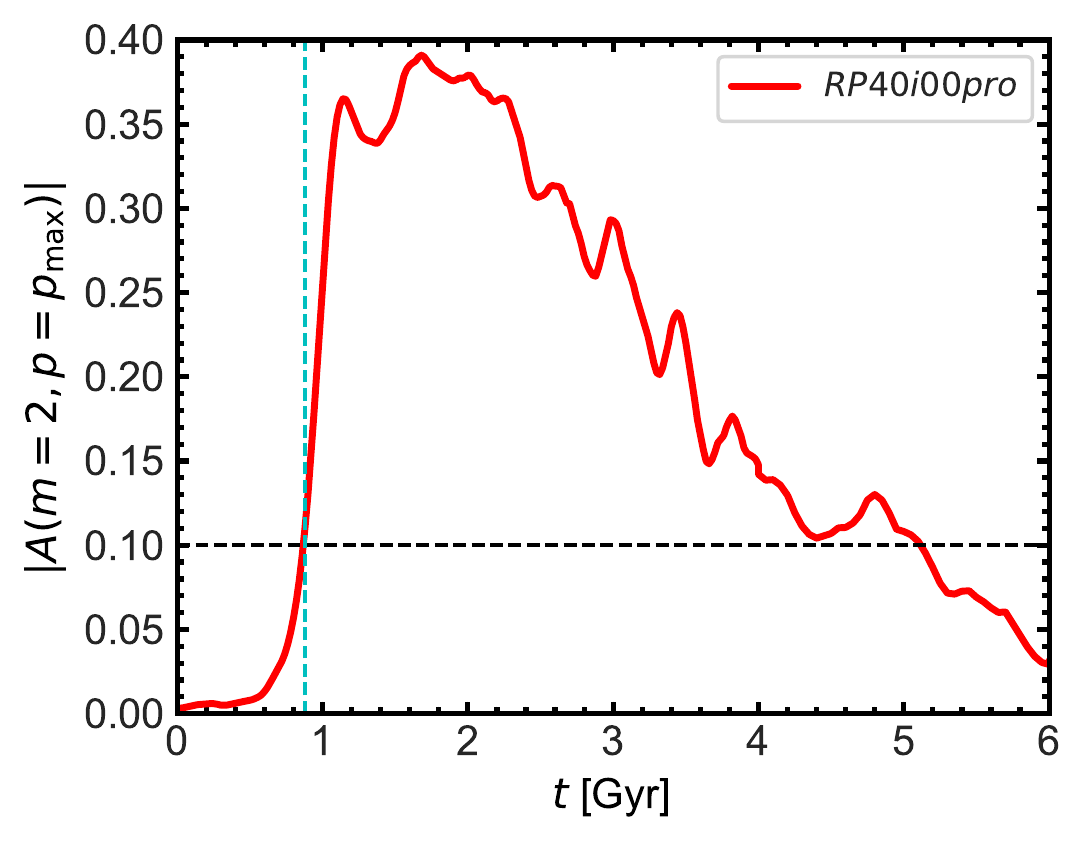}
    \caption{Temporal evolution of the spiral strength ($|A(m=2, p =p_{\rm max})|$), calculated using Eq.~\ref{eqn:spiral_strenght} for the model $RP40i00pro$. The tidally-induced spirals decay by the end of the simulation run. The horizontal black dotted line denotes  $|A(m=2, p =p_{\rm max})| = 0.1$, used as an operational definition for the onset of the spirals, see text for details. The vertical dashed line (in cyan) denotes the time of the pericentre passage of the perturber.}
    \label{fig:spiral_strength_temporal}
\end{figure}

Next, we quantify the temporal evolution of the strength of the tidally-induced spirals in our fly-by model $RP40i00pro$. Following \citet{Sellwood1984}, \citet{Sellwood1986}, \citet{Puerari.etal.2000}, we define
\begin{equation}
    A(m,p)=\frac{\sum_{j}^{N} m_{j} \exp[i(m\phi_{j}+p\ln R_{j})]}{\sum_{j}^{N} m_{j}},
    \label{eqn:spiral_strenght}
\end{equation}
\noindent where $|A(m, p)|$ is the amplitude, $m_{j}$ is the mass of $j^{th}$ star, $m$ is the spiral arm multiplicity, $(R_{j},\phi_{j})$ are the polar coordinates of the $j^{th}$ star in the plane of the disc, and $N$ is the total number of stellar particles in the annulus $R_{\rm min} \leq R \leq R_{\rm max}$ within which the spiral feature exists and/or is most prominent. Here, we take $R_{\rm min} = 3R_{\rm d}$, and $R_{\rm max} = 6 R_{\rm d}$, where $R_{\rm d} = 3.8 \kpc$. For this annular region, we estimate $A(m,p)$ as a function of $p$ for $p\in[-50, 50]$, with a fixed step of $dp=0.25$  \citep[as suggested by][]{Puerari.etal.2000} for different values of $m$. Then, we evaluate the parameter $p_{\rm max}$ which corresponds to the maximum value of $|A(m,p)|$.  We find that the amplitude $|A(m,p = p_{\rm max})|$ shows a maximum value for $m=2$,   indicating that the $m=2$ spiral is the strongest. Therefore, at a certain time $t$,  we define the amplitude $|A(m=2, p =p_{\rm max})|$ as the strength of spirals \citep{Sang-Hoon.etal.2015, Semczuk.etal.2017, Kumar.etal.2021}. The resulting temporal evolution of the strength of spirals for the model $RP40i00pro$ is shown in Fig.~\ref{fig:spiral_strength_temporal}. As seen from Fig.~\ref{fig:spiral_strength_temporal}, the tidally-induced spirals grow for some time after the interaction happens, then remain stable for about $1 \Gyr$ before weakening from around $t = 3 \Gyr$. By the end of the simulation, the spirals' strength becomes almost zero. For quantifying the longevity of the spirals, we define $|A(m=2, p =p_{\rm max})| = 0.1$ as the onset of the strong spiral perturbation\footnote{The threshold value of 0.1 is used purely as an operational definition for the onset of the spirals.}. Therefore, the spirals persist for a time of $\sim 4.2 \Gyr$ (after their formation) for the model $RP40i00pro$ (also see Fig.~\ref{fig:spiral_strength_temporal}).
\begin{figure}
    \centering
	\includegraphics[width=1.05\linewidth]{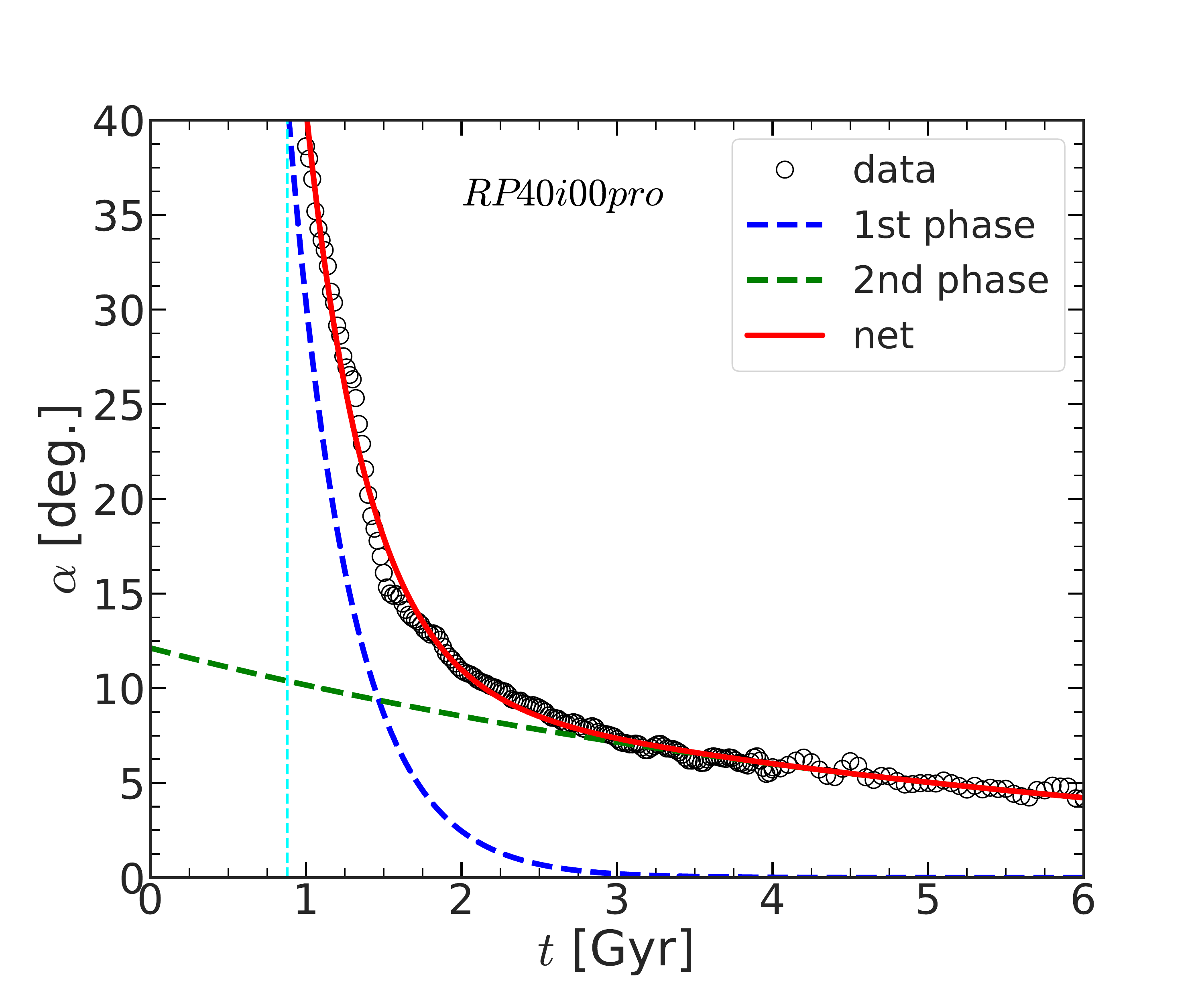}
    \caption{Temporal evolution of the pitch angle ($\alpha$) of the tidally-induced spirals is shown for the model $RP40i00pro$. A double-exponential profile (see Eq.~\ref{eq:winding_modelling}) is fitted to model the temporal evolution. The blue dashed line denotes the initial rapid winding phase whereas the green dashed line denotes the subsequent slow winding phase, for details see text. The red solid line denotes the best-fit double-exponential profile. The vertical dashed line (in cyan) denotes the time of the pericentre passage of the perturber.}
    \label{fig:winding_spiral}
\end{figure}

\subsection{Winding of the tidally-induced spirals}
\label{sec:winding_spiral}

Lastly, we investigate the winding of the tidally-induced spirals in our fly-by model $RP40i00pro$. As shown in past $N$-body simulations of galactic discs \citep[e.g. see][]{oh.etal.2008, Struck.etal.2011, Kumar.etal.2021}, a spiral arm can wind up with time. Following \citet{Sang-Hoon.etal.2015}, and \citet{Semczuk.etal.2017}, at a certain time $t$, we define the pitch angle, $\alpha$, as $\alpha=\tan^{-1}(m/p_{\rm max})$ using Eq.~\ref{eqn:spiral_strenght}. The resulting temporal evolution of the pitch angle for the model $RP40i00pro$ is shown in Fig.~\ref{fig:winding_spiral}. As revealed in Fig.~\ref{fig:winding_spiral}, the temporal evolution of the pitch angle displays two distinct phases, namely, the initial rapid winding phase where the pitch angle decreases sharply, and the subsequent slow winding phase where the pitch angle decreases less drastically. To model the temporal evolution of the pitch angle, we fit a double-exponential profile having the form
\begin{equation}
\alpha = \alpha_{1} \exp \left[-\lambda_1 (t - t_0) \right] + \alpha_{2} \exp \left[-\lambda_2 (t - t_0) \right]\,,
\label{eq:winding_modelling}
\end{equation}
\noindent where $\alpha_1$, $\alpha_2$, $\lambda_1$, and $\lambda_2$ are free parameters. Here, $t_0 = 1 \Gyr$, and denotes the time of the spirals' formation. The fitting is performed via the {\sc scipy} package {\sc curvefit} which uses the  Levenberg-Marquardt algorithm. The resulting best-fit double exponential profile is shown in Fig.~\ref{fig:winding_spiral}. We define the winding time-scale, $\tau_{\rm wind}$, as $\tau_{\rm wind} = \alpha/|\dot \alpha|$ where $|\dot \alpha|$ denotes the (absolute) rate of change of the pitch angle with time. For the initial rapid winding phase, $\tau_{\rm wind} \simeq 1/\lambda_1$ whereas for the subsequent slow winding phase $\tau_{\rm wind} \simeq 1/\lambda_2$ (see Eq.~\ref{eq:winding_modelling}). We find the best-fit values as $\lambda_1 =  2.51 \pm 0.05$ Gyr$^{-1}$ , and $\lambda_2 = 0.18 \pm 0.01$ Gyr$^{-1}$ which translate to a winding time-scale $\tau_{\rm wind} = 0.4 \Gyr$ for the initial rapid winding phase, and a winding time-scale $\tau_{\rm wind} = 5.7 \Gyr$ for the subsequent slow winding phase for the model $RP40i00pro$.

\subsection{Nature of tidally-induced spirals}
\label{sec:nature_spirals}

%
\begin{figure}
    \centering
	\includegraphics[width=\linewidth]{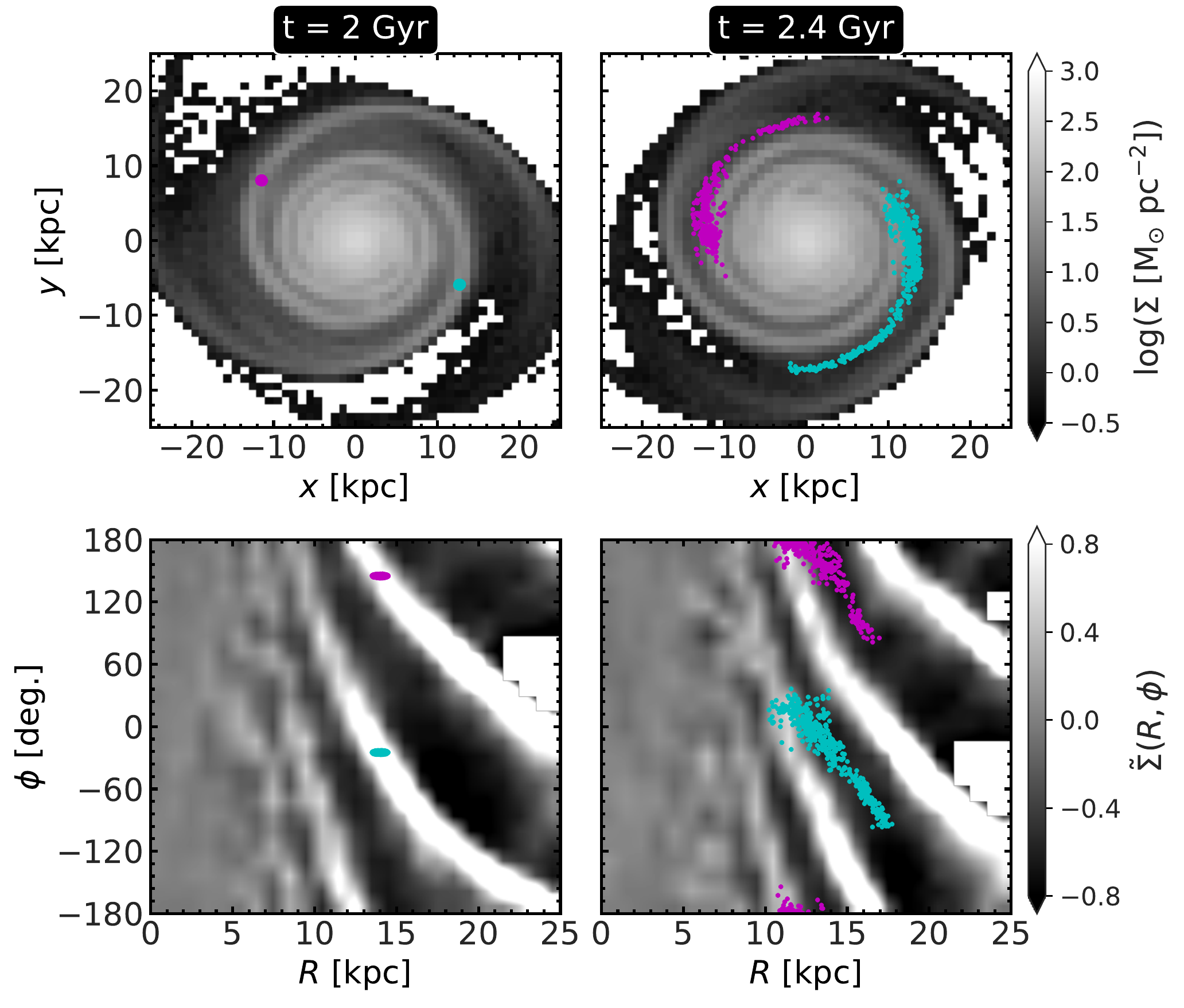}
    \caption{\textit{Nature of spirals:} top and bottom panels of the first column show the surface density and residual density maps of $RP40i00pro$ model at $t=2 \Gyr$ over-plotted with two patches (in magenta and cyan dots) of stellar particles in fully developed spiral arms. Right column is the analogous to the left column, but for $t=2.4 \Gyr$. Both patches of stars, initially associated along the spirals, have subsequently sheared out by the differential rotation, and left the spirals.}
    \label{fig:spiral_nature}
\end{figure}

So far, we have demonstrated that a fly-by interaction with a perturber induces a strong spiral feature in the outer disc of the host galaxy. However, the question remains whether these spirals are density waves or material arm in nature. For a comprehensive review on the nature of the spirals in disc galaxies, the reader is referred to \citet{BT08}, and \citet{DobbsandBaba2014}.
\par
For the set of the $N$-body models we are using here, we  do not have the age information of the stellar particles. This, in turn, restricts us from dividing stellar particles into different age-bins to trace the existence of spirals in different stellar population with different ages, as previously done in \citet{Ghosh.etal.2020}. Therefore, following \citet{Grand.etal.2012, D'Onghia.etal.2013}, we test the nature of the tidally-induced spirals by following the stars which are located on spirals arms at a certain time. We chose a time, say $t = 2 \Gyr$ when the spirals are fully-developed for the model $RP40i00pro$, and select two small patches (shown in magenta and cyan) of stars along the arms. This is shown in Fig.~\ref{fig:spiral_nature} (top left-hand panel). Now, if the spirals are of material arm in nature, then the stars would not leave the spiral arm at a subsequent time. To check that, we follow the selected stars at a later time, $t = 2.4 \Gyr$ (see top right-hand panel of Fig.~\ref{fig:spiral_nature}). As seen clearly from Fig.~\ref{fig:spiral_nature}, the stars, initially concentrated in small patches along the spiral arm, have sheared out due to the underlying differential rotation. This shearing out of the stars is more prominent in the distribution shown in the $(R, \phi)$-plane (see bottom panels of Fig.~\ref{fig:spiral_nature}). Interestingly, the stars, initially contained in the patches have left the spirals at a subsequent time, and the pitch angle of the selected stars is different from that of the spirals at $t = 2.4 \Gyr$. Furthermore, we calculate the pattern speed of the spirals at different radial locations in the disc. We find that the pattern speeds of the spirals are very different from the values of the circular frequency ($\Omega$), calculated at different radial locations. For brevity, these are not shown here. This shows that the spirals present in our fly-by model $RP40i00pro$ are density waves in nature.

\subsection{Dependence on orbital parameters}
\label{sec:orbital_variation}
%
\begin{figure}
    \centering
	\includegraphics[width=\columnwidth]{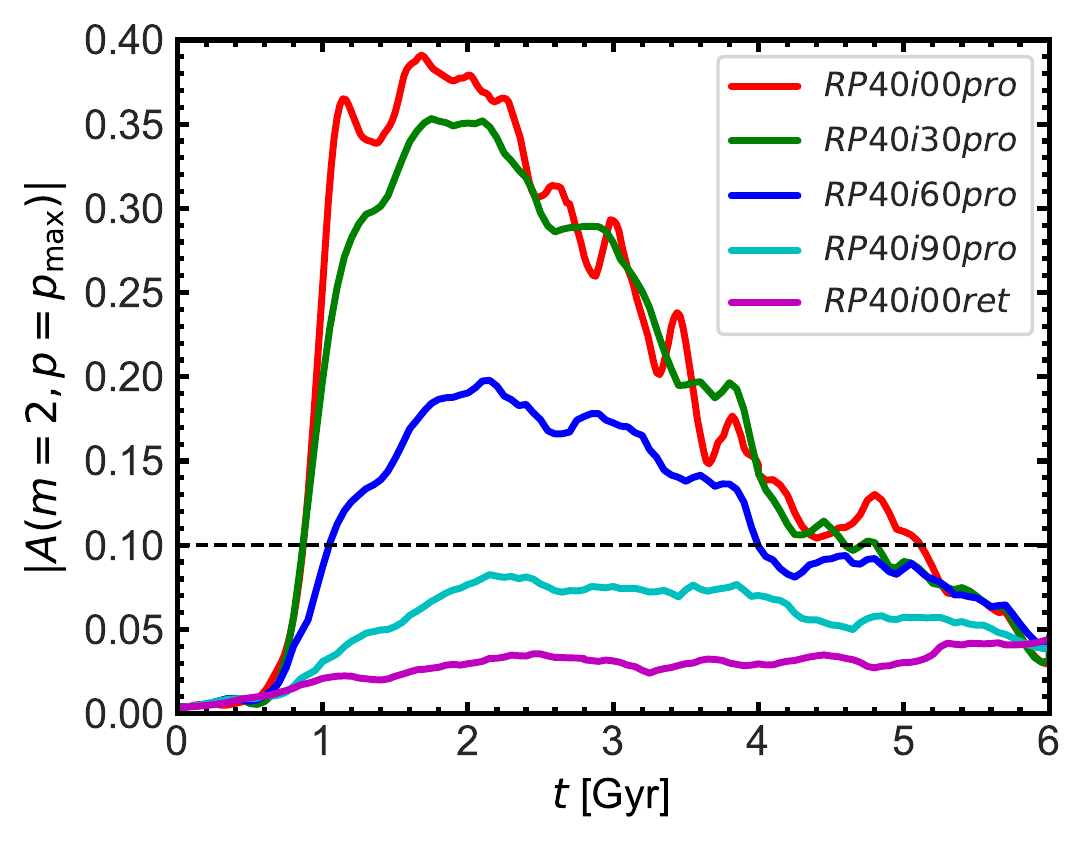}
    \caption{Temporal evolution of the spiral strength ($|A(m=2, p =p_{\rm max})|$), calculated using Eq.~\ref{eqn:spiral_strenght} is shown for models with different angle of interaction, and orbital spin vector. For comparison, we kept the model $RP40i00pro$ here (see red solid line). For models with prograde configuration, the strength of spirals decreases monotonically from the co-planar ($i =0 \degrees$) to polar ($i =90 \degrees$) orbital configuration. The horizontal black dotted line denotes  $|A(m=2, p =p_{\rm max})| = 0.1$, used as an operational definition for the onset of the spirals.}
    \label{fig:spiral_strength}
\end{figure}
Here, we explore the generation of the spirals in the host galaxy due to a tidal interaction with the perturber galaxy for different angles of interaction of the tidal encounter as well as for different orbital spin vectors (prograde or retrograde). Intuitively, the response of the host galaxy will be different for polar ($i =90 \degrees$) and co-planar ($i =0 \degrees$) orbital configurations. First we consider the other prograde models, with angles of interaction $30 \degrees$ and $60 \degrees$ (for details see section~\ref{sec:flyby_setup}). Both the models exhibit a similar trend of excitation of tidally-induced spirals in the disc of the host galaxy, as in model $RP40i00pro$. Shortly after the tidal encounter happens, the disc of the host develops a spiral feature which grows for a certain time, and then starts decaying. We calculate the strength of the spirals using Eq.~\ref{eqn:spiral_strenght} at different times for these models. The resulting temporal variations of the strength of the spirals for these three models are shown in Fig.~\ref{fig:spiral_strength}. The maximum strength of the tidally-induced spirals decreases monotonically with larger angle of interaction. Next, we use $|A(m=2, p =p_{\rm max})| = 0.1$ for defining the onset of the strong spiral perturbation. We find that the spirals persist for a time-scale of $\sim 2.9 - 4.2 \Gyr$, depending on the angle of interaction (also see Fig.~\ref{fig:spiral_strength}). For the polar ($i =90 \degrees$) configuration, the spirals are very weak as the values of $|A(m=2, p =p_{\rm max})|$ remain below $0.1$ (see Fig.~\ref{fig:spiral_strength}).

Next, we study the strength of the spirals in a galaxy fly-by model where the perturber interacts with the host galaxy in a retrograde orientation (with $i = 0 \degrees$) (for details see section~\ref{sec:flyby_setup}). A visual inspection of the face-on distribution of the stars (in the host galaxy) does not reveal any prominent spirals. We again calculate  the strength of the spirals using Eq.~\ref{eqn:spiral_strenght} for the model $RP40i00ret$, and show this also in Fig.~\ref{fig:spiral_strength}. The value of the amplitude $|A(m=2, p =p_{\rm max})|$ remains close to zero throughout the evolution of the model $RP40i00ret$, indicating that no prominent spirals are triggered/excited by the tidal encounter with the perturber galaxy for this model.
\par
Furthermore, we have checked that the tidally-induced spirals display a similar winding, namely, an initial rapid winding phase, followed by a slow winding phase, as seen for the model $RP40i00pro$ (see Fig.~\ref{fig:winding_spiral}). Also, we have checked the nature of the resulting spirals for these other models, using the same technique employed in section~\ref{sec:nature_spirals}. We find that for the models showing prominent spirals, the resulting spirals show a density wave nature, similar to the model $RP40i00pro$. For the sake of brevity, we do not show these results here.

\section{Breathing motions excited by tidally-induced spirals}
\label{sec:breathing}
%
\begin{figure}
    \centering
	\includegraphics[width=\columnwidth]{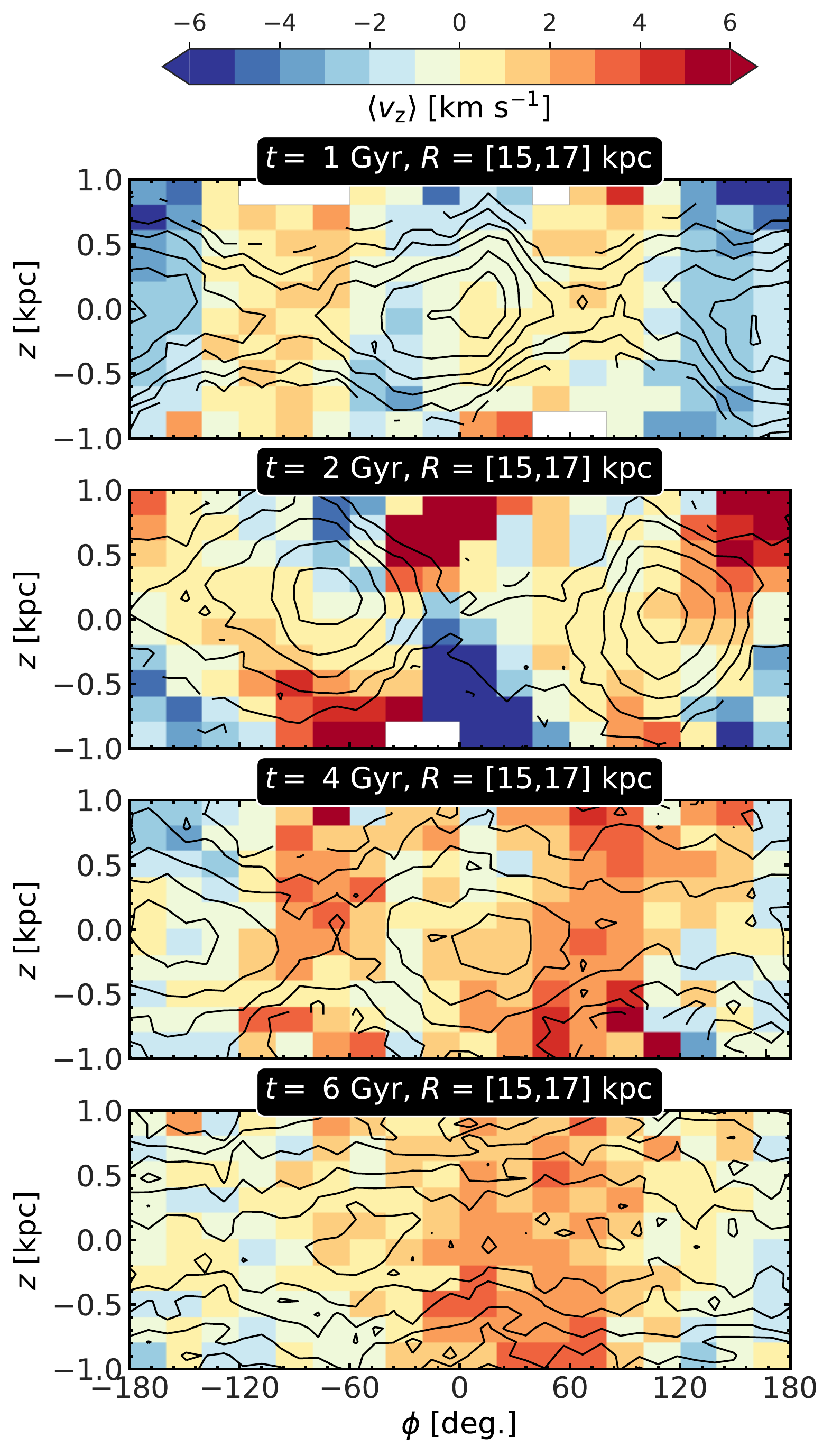}
    \caption{Distribution of the mean vertical velocity, $\avg{v_z}$, in the $(\phi, z)$-plane, for stars in the radial annulus $15 \leq R/\kpc \leq 17$ are shown at four different times for the model  $RP40i00pro$. The solid black lines denote the contours of constant density. The presence of large-scale, non-zero vertical velocities for stellar particles are seen at all four time-steps, for details see text.}
    \label{fig:vertical_velocity}
\end{figure}
\begin{figure*}
    \centering
	\includegraphics[width=\textwidth]{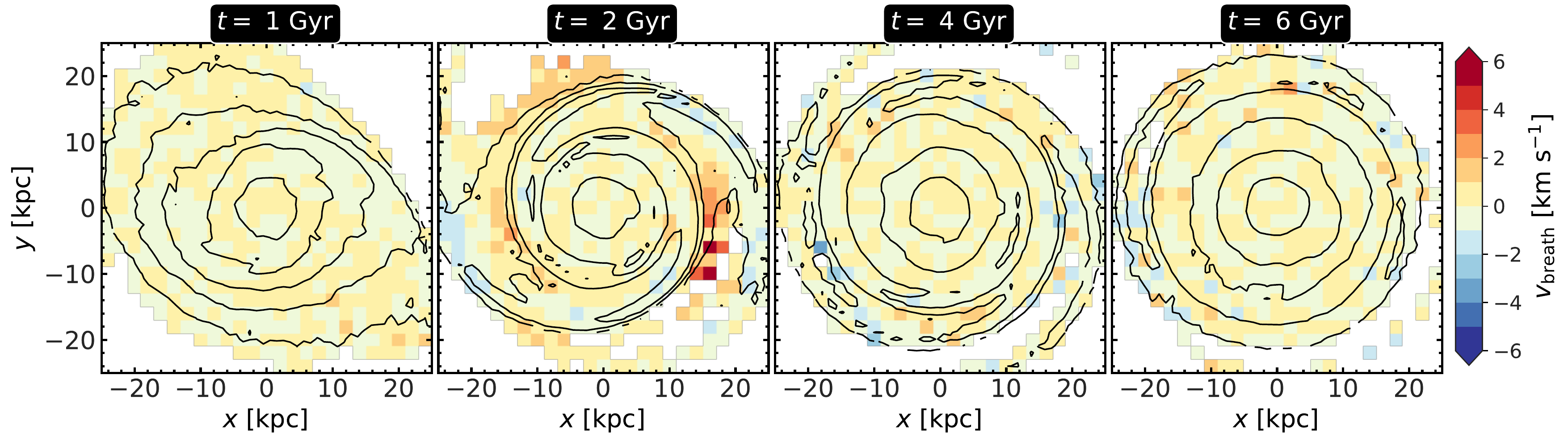}
	\medskip
    \includegraphics[width=\textwidth]{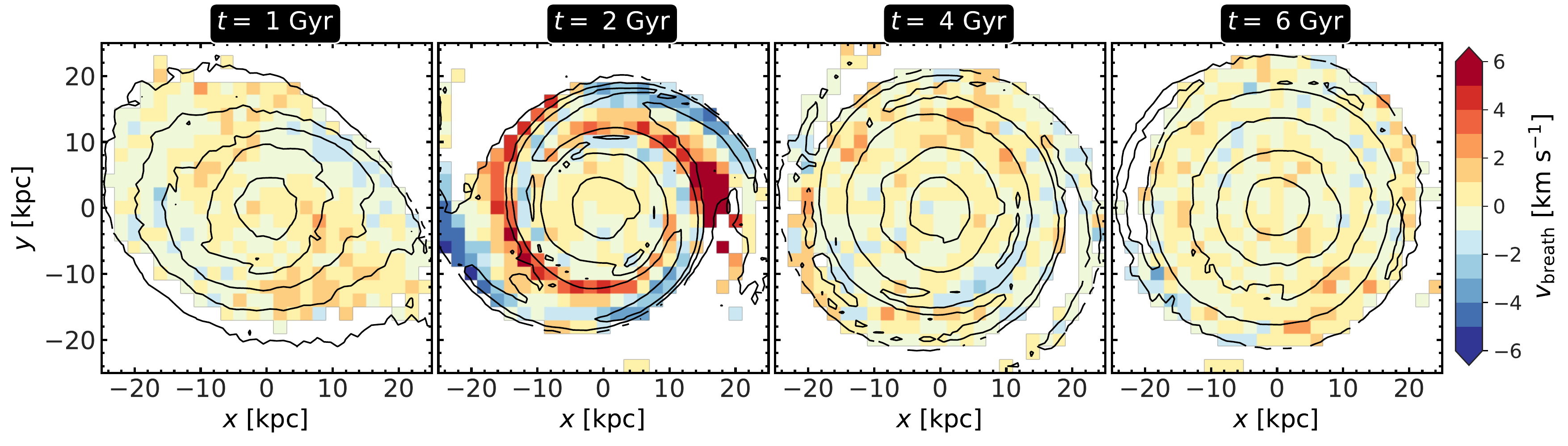}
    \caption{Distribution of the breathing velocity, \vb \ (Eq.~\ref{eq:breathing_velocity}) at different vertical distances from the mid-plane, at four different times for the model  $RP40i00pro$. The solid black lines denote the contours of constant density. The {\it top panels} show stars at $|z| = [0, 400]~\pc$ whereas the \textit{bottom panels} show stars at $|z| = [400, 1200]~\pc$, respectively.}
    \label{fig:breathing_map}
\end{figure*}
In the previous section, we have demonstrated that a tidal interaction with a perturber galaxy can excite prominent spirals in the disc of the host galaxy. Here, we investigate the dynamical impact of these tidally-induced spirals on the bulk vertical motions of the host galaxy. We first choose the fly-by model $RP40i00pro$ which harbours a strong spiral after the interaction. We choose the radial extent $15 \leq R/\kpc \leq 17$ where the spirals are prominent at later times in this radial annulus, and we calculate the mean vertical velocity ($\avg{v_z}$) in the $(\phi, z)$-plane at four different times, namely, at $t = 1, 2, 4,$ and $6 \Gyr$ (same as in Fig.~\ref{fig:dens_map}). This is shown is Fig.~\ref{fig:vertical_velocity}. During our chosen time interval, spirals' strength varies from strong to weak  (for details see section~\ref{sec:strength_spiral}).  At $t = 1 \Gyr$, just after the fly-by encounter, the distribution of the bulk vertical velocity ($\avg{v_z}$) in the $(\phi, z)$-plane predominantly shows bending motions, i.e., stellar particles on both sides of the mid-plane are moving coherently in the same direction. At this time, a prominent spiral is yet to form in the host galaxy (see Fig.~\ref{fig:spiral_strength_t-R}). However, by $t = 2 \Gyr$, the host galaxy shows a prominent spiral (see Fig.~\ref{fig:spiral_strength_t-R}), and the distribution of the bulk vertical velocity ($\avg{v_z}$) in the $(\phi, z)$-plane changes drastically. Now, the stellar particles on both sides of the mid-plane are moving coherently towards or away from it, indicating vertical breathing motion dominates. The relative sense of the $\avg{v_z}$ varies as a function of the azimuthal angle. However, at $t = 4 \Gyr$ when the tidally-induced spirals have weakened substantially, the distribution of the $\avg{v_z}$ is again seen to be dominated by the bending motions. By the end of the simulation  ($t = 6 \Gyr$), the spiral has wound up, and the distribution of $\avg{v_z}$ remains dominated by the bending motions of the stars.
\par
To quantify the breathing motions, we define the breathing velocity, $\vb $, as \citep{Debattista.2014, Gaia.Collaboration.2018, Ghosh.etal.2020}
\begin{equation}
\vb(x, y) = \frac{1}{2}\left[\avg{v_z(x,y, \Delta z)} - \avg{v_z(x,y, -\Delta z)}\right]\,,
    \label{eq:breathing_velocity}
\end{equation}
\noindent where $\avg{v_z(x,y, \Delta z)}$ is the mean vertical velocity at position $(x,y)$ in the galactocentric cartesian coordinate system, averaged over a vertical layer of thickness $\Delta z$ \citep[for details see][]{Gaia.Collaboration.2018, Ghosh.etal.2020}. A positive breathing velocity ($\vb > 0$) implies that the stars are coherently moving away from the mid-plane (expanding breathing motion), while $\vb < 0$ implies that the stars are moving coherently towards the mid-plane (compressing breathing motion). We find that at larger heights, the particle resolution of our selected fly-by model is not well-suited to compute meaningful values of $\vb$. Therefore, we calculate the distribution of $\vb$ in the $(x,y)$-plane for two vertical layers, namely, $|z| = [0, 400] \pc$ and $|z| = [400, 1200] \pc$.
The resulting distributions of $\vb$ at the same four times (as in Fig.~\ref{fig:vertical_velocity}) are shown in Fig.~\ref{fig:breathing_map}. The breathing velocity is close to zero near the mid-plane, for all four times considered. However, at $t = 2 \Gyr$ when the strong spirals are present in the host galaxy, the upper vertical layer ($|z| = [400, 1200] \pc$) shows significant, coherent breathing velocity ($\sim 6-8 \kms$, in magnitude). This trend is similar to what is shown for spiral-driven breathing motions \citep[e.g., see][]{Debattista.2014,Ghosh.etal.2020}, and is also similar to the breathing motions seen from the \gaia\ DR2 \citep{ Gaia.Collaboration.2018}. A visual inspection also reveals that at $t = 2 \Gyr$, the compressing breathing motions are closely associated with the spiral arms whereas the expanding breathing motions arise in the inter-arm regions (see Fig.~\ref{fig:breathing_map}), similar to the findings of \citet{Faureetal2014}, \citet{Debattista.2014} and \citet{Ghosh.etal.2020}. However, after $t = 4 \Gyr$ when the spirals are either significantly weaker or completely absent, the host galaxy does not show any prominent, coherent breathing motion in the upper layer ($|z| = [400, 1200] \pc$). To probe this further, we calculate the distribution of $\vb$ at two vertical layers in every $200 \Myr$ from $t = 2 \Gyr$ to $t = 6 \Gyr$ for the model  $RP40i00pro$. For the sake of brevity, this is not shown here. We find that the incidence of a prominent, coherent breathing motion is strongly related with the presence of a strong spiral feature in the host galaxy. This, together with the amplitude of the breathing velocity increasing with height \citep[a signature of spiral-driven breathing motion, see][]{Ghosh.etal.2020} indicate that the breathing motions present in the host galaxy are driven by the tidally-induced spirals.
\begin{figure}
    \centering
	\includegraphics[width=\linewidth]{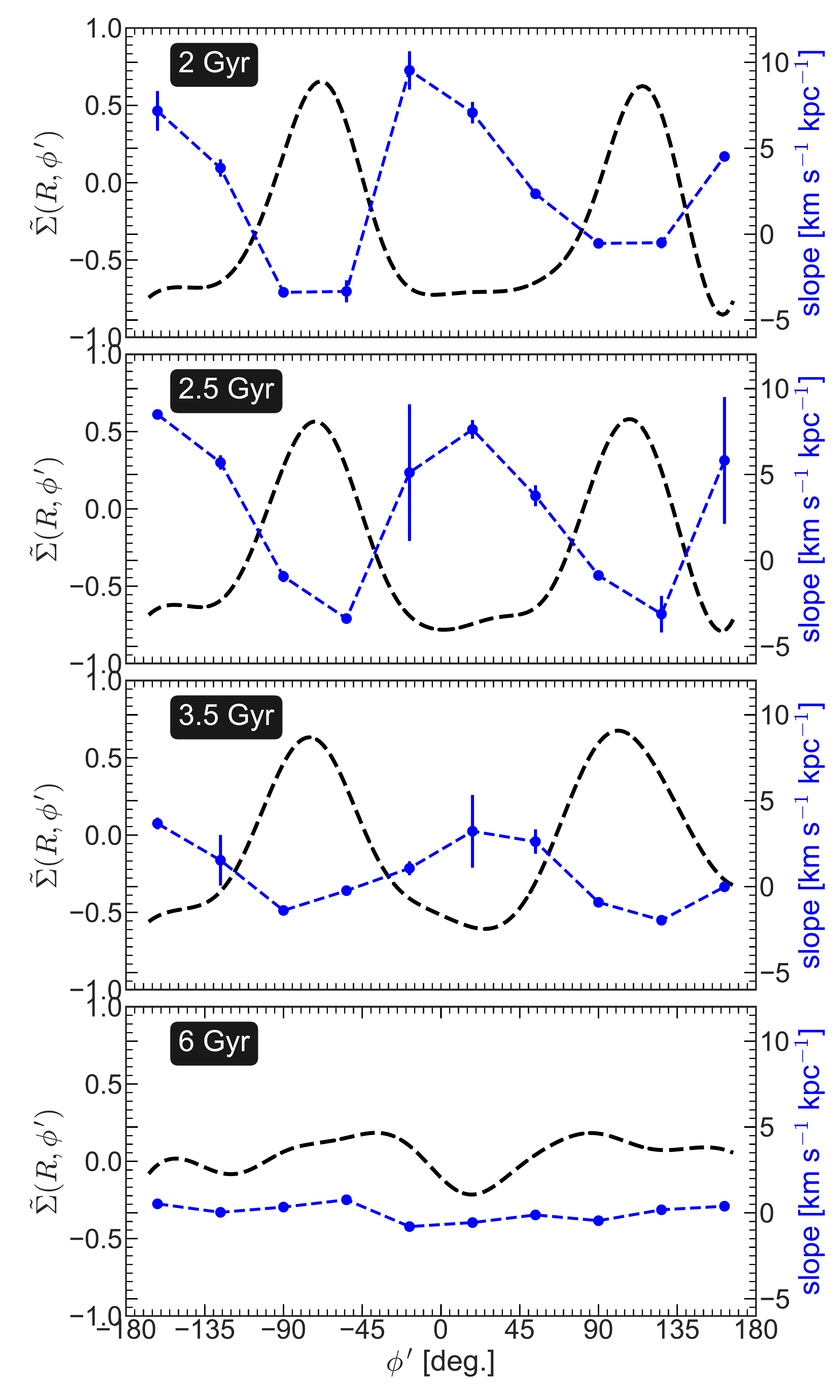}
    \caption{Variation of the slopes as a function of the rotated azimuthal angle ($\phi'$) at four different times (having different spiral strength) for the model $RP40i00pro$ (see blue dashed lines). Only the stellar particles in the radial extent $15 \leq R/\kpc \leq17 $ are chosen here. The black dashed line denotes the residual surface density ($\tilde \Sigma (R, \phi')$) as a function of the azimuthal angle, calculated in this chosen radial extent. The particles have first been binned in $1~\kpc$ annuli and then azimuthally rotated relative to each other so that the minimum in $\tilde \Sigma (R, \phi)$) in each annulus is coincident, and then the mean vertical velocity ($\avg{v_z}$) is calculated, amounting to stacking different radial ranges while unwinding the spiral. For details, see text. The sense of rotation is towards increasing $\phi'$.}
\label{fig:meanvz_resden_azimuthal_variation}
\end{figure}
\par
Finally, we study the azimuthal variation of the breathing motions and their connection with the peak(s) of the spirals. Figs.~\ref{fig:vertical_velocity} and \ref{fig:breathing_map} already demonstrated that the breathing motions in model  $RP40i00pro$, is associated with the spiral arm and the inter-arm regions. Here, we explore this further. Following \citet{Ghosh.etal.2020}, we quantify these breathing motions whose amplitude increases with height from the mid-plane, by fitting a straight line. In this formalism, the presence of a prominent breathing motion would result in a significantly \textit{non-zero} slope. Furthermore, an expanding (positive) breathing motion will yield a positive slope whereas a compressing (negative) breathing motion will yield a negative slope. On the other hand, the non-zero value of the intercept of the best-fit straight line indicates the presence of a bending motion. For details, the reader is referred to \citet{Ghosh.etal.2020}. A similar approach was also used in \citet{Widrowetal2014}. We consider the same radial extent $15 \kpc \leq R \leq 17 \kpc$ where a prominent spiral is present at $t = 2 \Gyr$. Since the phase-angle of the $m=2$ Fourier mode ($\phi_2$) varies as a function of radius, indicating that the azimuthal locations of the density peaks vary as a function of radius, to obtain a stronger signal of the slope, we first rotate the stellar particles in two different $1\kpc$-wide radial bins, in such a way that the density peaks in our chosen radial extent coincide. Then, we recalculate the slope of the breathing velocity as a function of the rotated azimuthal angle ($\phi'$). The resulting variation of the slope as well as the residual surface density ($\tilde \Sigma(R, \phi')$) with rotated azimuthal angle ($\phi'$) are shown in Fig.~\ref{fig:meanvz_resden_azimuthal_variation} for four different times. As seen clearly from Fig.~\ref{fig:meanvz_resden_azimuthal_variation}, at $t = 2 \Gyr$ when the spirals are quite strong ($|A(m=2, p = p_{\rm max})| = 0.38$), the compressive breathing motions are associated with the peak(s) of the spiral whereas the expanding  breathing motions are associated with the density minima. This trend is consistent with the signature of a spiral-driven breathing motions as shown in \citet{Debattista.2014} and \citet{Ghosh.etal.2020}. We show that in our model, the amplitudes of the expanding breathing motions are higher than that of the compressing breathing motions -- a trend attributed to the fact of having a more abrupt density variation as the stellar particles leave the spirals compared with when they enter the spirals \citep[see][]{Debattista.2014}. At $t = 3.5 \Gyr$, when the spirals are decaying ($|A(m=2, p = p_{\rm max})| = 0.21$), the corresponding (absolute) values of the slope also decrease, thereby indicating that the breathing motions also weaken. By $t = 6 \Gyr$, the spirals get wound up (almost) completely ($|A(m=2, p = p_{\rm max})| = 0.04$), and the (absolute) values of the slope are close to zero, indicating the absence of breathing motion. This further strengthens the case for the breathing motions being driven by the spirals, and not the \textit{direct} dynamical consequence of a fly-by interaction. The breathing motions are seen to last for $\sim 1.5 -2 \Gyr$ after their generation for the model $RP40i00pro$.
\par
We repeated this whole set of analyses for all the other models, to investigate the breathing motions and their connection with the incidence and strength of spirals present in the model. We find that a strong spiral always drives a prominent breathing motion with expanding breathing motions associated with the inter-arm region whereas the compressing breathing motions are associated with the peak(s) of the spirals. As the spirals get wound up, the breathing motions also cease to exist in the fly-by models. These trends are similar to what we have found for the model $RP40i00pro$. Therefore, for brevity, these are not shown here. In the fly-by model with retrograde orbital configuration ($RP40i00ret$), the spiral structure itself is weaker compared to the other models in prograde orbital configuration (see Fig.~\ref{fig:spiral_strength}). We checked that no prominent breathing motion is excited by this feeble spiral structure.

\section{Discussion}
\label{sec:discussion}
Our fly-by interactions excite strong spirals in the outer regions of the host galaxy's disc. The spirals show a variation in their maximum strength depending on the angle of interaction, and the orbital spin vector. Here, we compare the  strength, location, and nature of the tidally-induced spirals in our models with past studies. The numerical simulations of \citet{Oh.etal.2015} showed that in their models, a stronger tidal encounter induces prominent spirals in the inner regions ($ 5 \leq R/\kpc \leq 10$) of the host galaxy's disc (in addition to the tidal tails in the outer parts). The arm strength of the spirals vary in the range $\sim 0.1 - 0.18$, depending on the values of the relative tidal force, and the relative imparted momentum. Moreover, the spirals are of kinematic density wave in nature. Also, the simulations of \citet{Semczuk.etal.2017} showed the generation of transient spirals due to the tidal force exerted by the potential of a cluster; the spirals appear with each pericentre passage, followed by a fast decay. These spirals appear in the outer disc region ($ 12 \leq R/\kpc \leq 17$), with the maximum arm strength varying in the range $\sim 0.4 - 0.75$. In comparison, the spirals in our fly-by models are most prominent in the outer regions ($R \geq 10 \kpc$). The maximum arm strength varies in the range $\sim 0.15-0.38$, depending on the angle of interaction, and the orbital spin vector (see Fig.~\ref{fig:spiral_strength}). The spirals appear shortly after the pericentre passage of the perturber, grow rapidly, followed by winding up of those spirals. This winding of spirals show two distinct phases, namely, the initial rapid winding phase, followed by a slow winding phase (see Fig.~\ref{fig:winding_spiral}).
\par
The amplitude of the spiral-driven breathing motions also merits some discussion. In our fly-by models, when the spirals are most prominent, the values of the best-fit slope (used as a proxy for the breathing amplitude) vary from $\sim -5 \kms$ kpc$^{-1}$ to $\sim 10 \kms$ kpc$^{-1}$ (see Fig.~\ref{fig:meanvz_resden_azimuthal_variation}).  In comparison, the fly-by model (with spirals present) of \citet{Widrowetal2014} showed the (absolute) values of the best-fit slope $\sim 10 \kms$  kpc$^{-1}$. As for the spiral-driven breathing motions where the spirals arise due to internal instability, \citet{Ghosh.etal.2020} reported the values of the slope varying from $\sim - 2.5 \kms$ kpc$^{-1}$ to $\sim 3 \kms$ kpc$^{-1}$. Furthermore, the strong spirals present in the models used by \citet{Faureetal2014,Debattista.2014} can drive large breathing motions ($|\avg{v_z}|~ \sim 5-20 \kms$).
\par
We have considered only $N$-body models of an unbound, fly-by interaction, excluding the interstellar gas. It is well known that a disc galaxy contains a finite amount of gas \citep[e.g., see][]{ScovilleandSanders1987}. Additionally, in the $\Lambda$CDM galaxy formation scenario, a galaxy can accrete cold gas \citep[e.g.,][]{BirnboimDekel2003,Keresetal2005,DekelBirnboim2006,Ocvriketal2008} either during the merger-phase or at a later stage. Past studies have shown the dynamical importance of the interstellar gas in the context of cooling the stellar disc and facilitating the generation of fresh spiral waves \citep{SellwoodCarlberg1984}, and in maintenance of spiral density waves in infinitesimally-thin discs \citep{GhoshJog2015,GhoshJog2016} as well as in a galactic disc with finite thickness \citep{GhoshJog2021}. For gas rich galaxies undergoing such fly-by interactions, the vertical breathing motion may be important for increasing the turbulence in the gas where star formation is insignificant \citep{Stilp.etal.2013}. This is because as the stars are in the breathing motion, the vertical potential will change with time and the gas distribution will be affected. However, since the vertical stellar velocity induced by the breathing motion is fairly small (of the order of a few km s$^{-1}$), this effect may not be very significant, especially if compared to the much larger kinematic effect of supernova explosions and stellar winds \citep{Krumholz.etal.2018, Yu.Brian.etal.2021}. In addition, galaxy bulges play an important role in maintaining spiral density waves in the disc for a longer time \citep{SahaandElmegreen2016}. Although, our galaxy models have a classical bulge, we have not varied the contribution of the bulges in our models.
\par
Lastly, our models are specifically designed in a way that the unperturbed disc galaxy is not bar unstable and only forms weak spirals; the strong spirals that form therefore are a dynamical result of the fly-by encounter.
However, in reality, the Milky Way also harbours a stellar bar \citep[e.g., see][]{LisztandBurton1980,Binneyetal1991,Weinberg1992,Binneyetal1997,BlitzandSpergel1991,Weiland.etal.1994,Dwek.etal.1995,Freudenreich.1998,Hammersleyetal2000,WegandGerhard2013}. Furthermore, the bulk vertical motions in the Solar neighbourhood and beyond, display both bending and breathing motions \citep[e.g.,][]{Gaia.Collaboration.2018,Carrilloetal2018}. This simultaneous presence of bending and breathing motions could well be collectively manifested due to a combination of internal (spiral and/or bar-driven) and external driving mechanisms (tidal encounters), as previously investigated by \citet{Carrilloetal2019}. We stress that the aim of this work is to clarify whether the excitation of breathing motions are `directly' related to tidal interactions or whether they are driven by the tidally-induced spirals (as also mentioned in section~\ref{sec:intro}), and not to replicate the observed dynamical state of the Milky Way.

\section{Summary }
\label{sec:conclusion}
In summary, we investigated the dynamical impact of an unbound, single fly-by interaction with a perturber galaxy on the generation of the tidally-induced spiral features and the associated excitation of vertical breathing motions. We constructed a set of $N$-body models of fly-by encounter, with mass ratio kept fixed to 5:1 while varying different orbital parameters.
Our main findings are :\\

\begin{itemize}

\item{Fly-by interactions trigger a strong spiral structure in the disc of the host galaxy. The spirals grow rapidly in the initial times, followed by a slow decay. The generation and the strength of these tidally-induced spirals depend strongly on the angle of interaction as well as on the orbital spin vector. For the same orbital energy and the angle of interaction, the models in prograde configuration are  more efficient at driving strong spirals when compared to models in retrograde configuration.}

\item{The tidally-induced spirals in the host galaxy can survive for $\sim 2.9 - 4.2 \Gyr$ after their formation. The pitch angle of the resulting spirals display two distinct phases of winding, namely, a fast winding phase ($\tau_{\rm wind} \sim 0.4 \Gyr$) and a subsequent slow winding phase ($\tau_{\rm wind} \sim 5.7 \Gyr$). }

\item{When the tidally-induced spirals are strong, they drive coherent, large-scale vertical breathing motions whose amplitude increases with height from the mid-plane. Furthermore, the azimuthal locations of the compressing breathing motions ($\vb <0$) are associated with the peaks of the spirals whereas the azimuthal locations of the expanding breathing motions ($\vb > 0$) coincide with the density minima of the spirals. These trends are in agreement with the signatures of spiral-driven breathing motions.}

\item{The temporal evolution of these breathing motions follow closely the temporal evolution of the strengths of the spirals. A stronger spiral drives breathing motions with larger amplitudes. These breathing motions, excited by tidally-induced spirals, can persist for $\sim 1.5-2 \Gyr$ in the disc of the host galaxy.}

\end{itemize}

Thus, the results presented in this paper demonstrate that a strong spiral structure can drive large, coherent vertical breathing motions irrespective of their formation scenario, i.e., whether induced by tidal interactions (as shown here) or generated via internal disc gravitational instability \citep[e.g.,][]{Faureetal2014,Debattista.2014,Ghosh.etal.2020}. Furthermore, our results highlight the cautionary fact that although in past studies, the tidal interactions are considered as the `usual suspect' for driving the vertical breathing motions, it is indeed the tidally-induced spirals which drive the breathing motions, and the dynamical role of such tidal encounters remains only ancillary.

\section*{ACKNOWLEDGEMENTS}
We thank the referee, Matthias Steinmetz, for valuable suggestions that have improved the paper. We are grateful to the high performance computing facility `NOVA' at the Indian Institute of Astrophysics, Bengaluru, India, and the facilities of the Center for High Performance Computing at Shanghai Astronomical Observatory, Shanghai, China where we ran all our simulations. A.K. thanks Denis Yurin and Volker Springel for providing GalIC and Gadget-2 code that is used for this study. This study also made use of NumPy \citep{numpy2020}, Matplotlib \citep{matplotlib2007}, and Astropy \citep{astropy2018} packages. M.D. acknowledges the support of Science and Engineering Research Board (SERB) MATRICS grant MTR/2020/000266 for this research. 

\section*{DATA AVAILABILITY}
The data generated in this research will be shared on reasonable request to the corresponding author.




\bibliographystyle{mnras}
\bibliography{Kumar} 



\appendix

\section{Evolution in isolation}
\begin{figure*}
    \centering
	\includegraphics[width=0.6\textwidth]{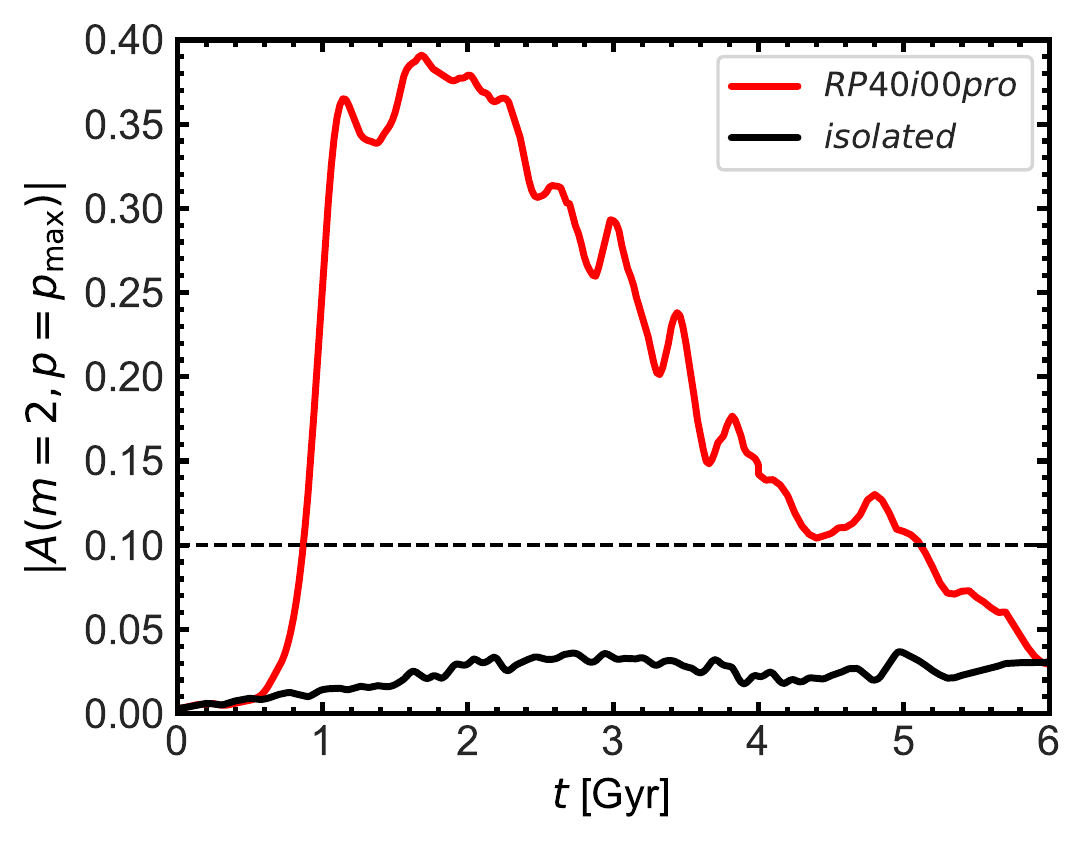}
    \caption{Temporal evolution of the spiral strength ($|A(m=2, p =p_{\rm max})|$), calculated using Eq.~\ref{eqn:spiral_strenght} is shown for the isolated host galaxy model. For comparison, we kept the model $RP40i00pro$ here (see red solid line). The isolated model does not show any prominent spirals throughout the simulation run. }
    \label{fig:spiral_strength_isolated}
\end{figure*}

\begin{figure*}
    \centering
	\includegraphics[width=\textwidth]{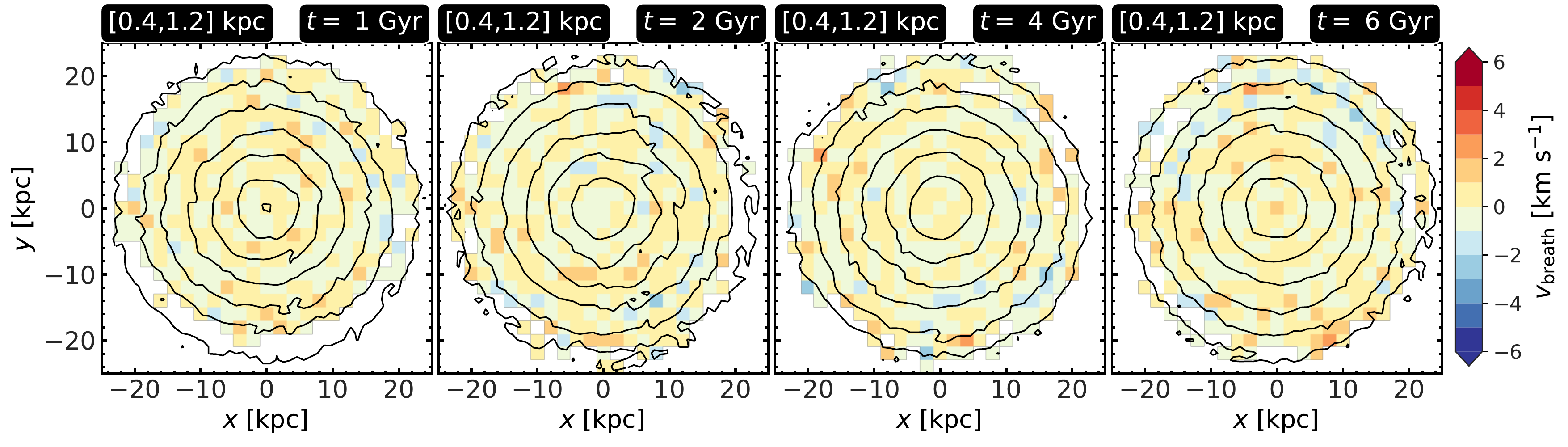}
    \caption{Evolution in isolation: distribution of the breathing velocity, \vb, for stars at $|z| = [400, 1200]~\pc$ at four different times for the isolated host galaxy model. The solid black lines denote the contours of constant density. No prominent vertical breathing motion is seen for the isolated evolution, in sharp contrast with the fly-by models (e.g., compare with Fig.~\ref{fig:breathing_map}).}
    \label{fig:breathing_map_isolated}
\end{figure*}

So far, we have shown that in our fly-models, a prominent spiral appears shortly after the interaction happens, and this spiral drives a coherent vertical breathing motion in the host galaxy. However, it remains to be investigated whether the host galaxy, when evolved in isolation, could still generate spirals and the associated vertical breathing motions.  We evolve the host galaxy model in isolation, for $6 \Gyr$. A visual inspection of the face-on density distribution of the stellar particles reveals no prominent spirals, throughout the simulation run, (as can be seen in the density contours in Fig.~\ref{fig:breathing_map_isolated}). Following the methodology described in Section~\ref{sec:strength_spiral}, we measure the values of $|A(m=2, p =p_{\rm max})|$ (used to quantify the strength of spirals) for the isolated host galaxy model. The resulting temporal evolution is shown in Fig.~\ref{fig:spiral_strength_isolated}. The values of $|A(m=2, p =p_{\rm max})|$ remain close to zero throughout the simulation run, demonstrating that no prominent spiral arm is generated during the entire isolated evolution of the host galaxy model. 

Furthermore, we calculate the breathing motions ($\vb$), using Eq.~\ref{eq:breathing_velocity}, for both the vertical slices, namely, $|z| = [0, 400] \pc$ and $|z| = [400, 1200] \pc$. As before, the vertical slice $|z| = [0, 400] \pc$ does not show any breathing motion. Interestingly, the upper vertical slice ($|z| = [400, 1200] \pc$) does not show any prominent breathing motion either throughout the simulation run (see Fig.~\ref{fig:breathing_map_isolated}), in sharp contrast with the fly-by models (compare with Fig.~\ref{fig:breathing_map}). This clearly demonstrates that the spirals in the fly-by models are indeed tidally-induced. In other words, the generation of spirals and the associated spiral-driven vertical breathing motions can indeed be attributed to the dynamical impact of a fly-by interaction.

%


\bsp	
\label{lastpage}
\end{document}